\def\spose#1{\hbox to 0pt{#1\hss}}
\def\simlt{\mathrel{\spose{\lower 3pt\hbox{$\mathchar"218$}}
     \raise 2.0pt\hbox{$\mathchar"13C$}}}
\def\simgt{\mathrel{\spose{\lower 3pt\hbox{$\mathchar"218$}}
     \raise 2.0pt\hbox{$\mathchar"13E$}}}
\def\simpropto{\mathrel{\spose{\lower 3pt\hbox{$\mathchar"218$}}
     \raise 2.0pt\hbox{$\propto$}}}

\documentclass[useAMS,usenatbib]{mn2e}
\usepackage{amsmath,graphicx,bm}
\title{Will point sources spoil 21 cm tomography?}
\author[Adrian Liu, Max Tegmark, Matias Zaldarriaga]{Adrian Liu$^{1}$\thanks{E-mail:
acliu@mit.edu}, Max Tegmark$^{1}$, Matias Zaldarriaga$^{2,3}$\\
$^{1}$Dept. of Physics and MIT Kavli Institute, Massachusetts Institute of Technology, 77 Massachusetts Ave., Cambridge, MA 02139, USA\\
$^{2}$Harvard-Smithsonian Center for Astrophysics, 60 Garden St., Cambridge, MA 02138, USA\\
$^{3}$Jefferson Laboratory of Physics, Harvard University, Cambridge, MA 02138, USA}
\date{Submitted to {\it MNRAS} July 20, 2008 }
\begin{document}

\pagerange{\pageref{firstpage}--\pageref{lastpage}} \pubyear{2008}

\maketitle

\begin{abstract}
$21$~cm tomography is emerging as a promising probe of the cosmological dark ages and the epoch of reionization, as well as a tool for observational cosmology in general.  However, serious sources of foreground contamination must be subtracted for experimental efforts to be viable. In this paper, we focus on the removal of unresolved extragalactic point sources with smooth spectra, and evaluate how the residual foreground contamination after cleaning depends on instrumental and algorithmic parameters. A crucial but often ignored complication is that the synthesized beam of an interferometer array shrinks towards higher frequency, causing complicated frequency structure in each sky 
pixel as ``frizz'' far from the beam centre contracts across unresolved radio sources. We find that current-generation experiments should nonetheless be able to clean out this points source contamination adequately, and quantify the instrumental and algorithmic design specifications required to meet this foreground challenge.

\end{abstract}


\section{Introduction}
The use of the $21$ cm line of neutral hydrogen as a probe of cosmological structure has generated much excitement over the last few years. $21$ cm tomography promises to place extremely tight constraints on cosmological parameters \citep{Matt3,Santos2,Wyithe, juddjackiemiguel1,Yi}, as well as to act as a direct probe of the cosmological dark ages and the Epoch of Reionization (EOR) \citep{Rees, Tozzi2, Tozzi, Iliev,   furlanetto1, Loeb1,furlanetto2, Barkana1,Mack}.  As a result, many $21$~cm experiments in the form of large radio arrays have recently been funded and are currently being built, e.g., LOFAR\footnote{http://www.lofar.org}, 21CMA\footnote{http://21cma.bao.ac.cn/}, MWA\footnote{http://www.haystack.mit.edu/ast/arrays/mwa/} and PAPER\footnote{http://astro.berkeley.edu/~dbacker/eor/}.

\begin{figure}
\centering
\includegraphics[width=0.5\textwidth]{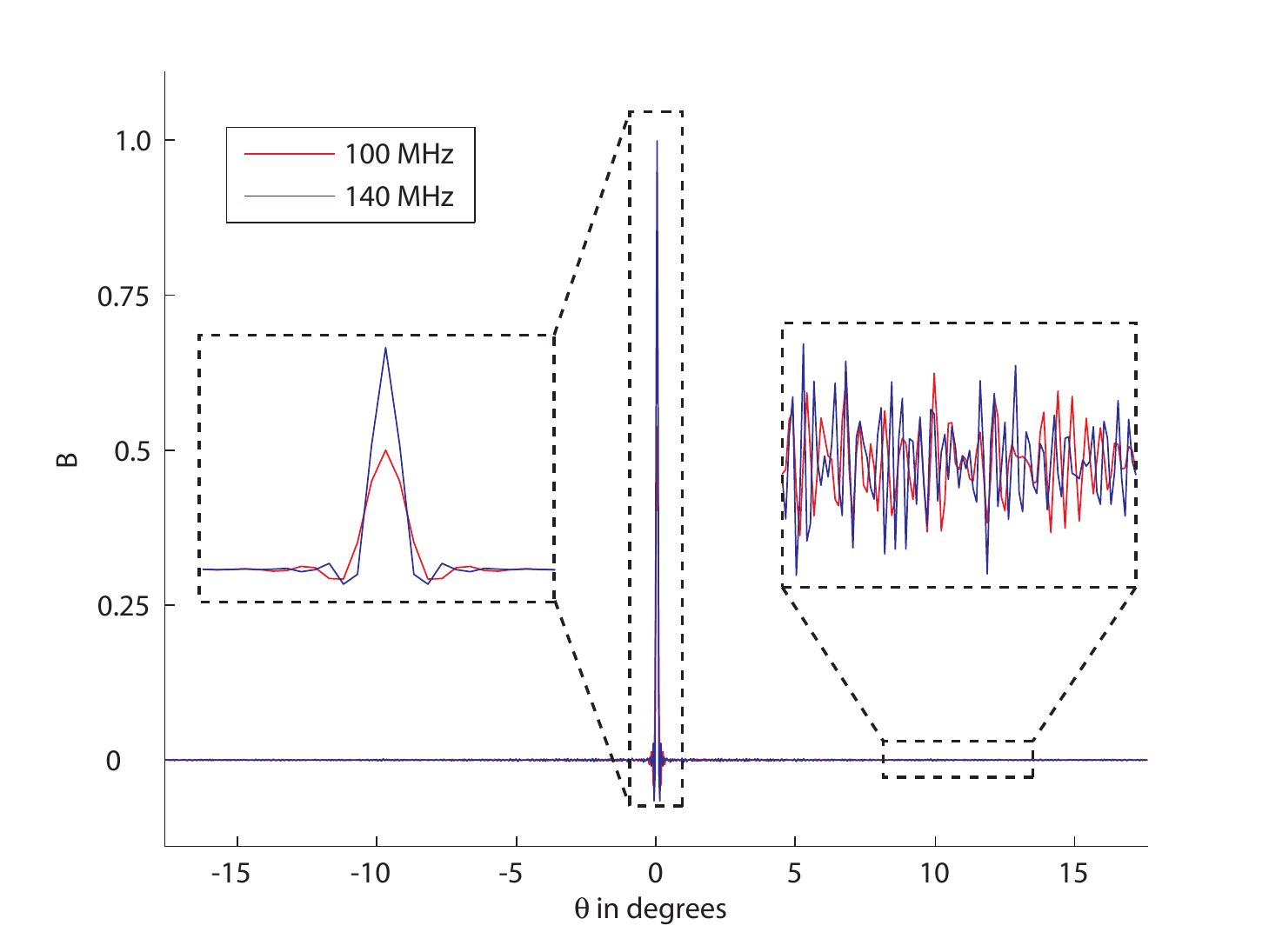}
\caption{Sample synthesized beam profile for a typical 21-cm tomography experiment with 500 antenna tiles.  The tiles are distributed within a diameter $D_{\rm array} \sim 1500\,\textrm{m}$ according to the density function $\rho (r) \sim r^{-2}$.  From the profile, it is clear that a realistic evaluation of foreground subtraction techniques should take into account the fact that beam widths vary as $\lambda /D$.  This is particularly important when subtracting off unresolved point sources, because a point source can fall on an off-beam ``spike" that has an effectively unpredictable dependence on frequency.}
\label{awesome}
\end{figure}

The observational challenges, however, are daunting.  At radio frequencies, one must deal with a long list of contaminants that have the ability to swamp the cosmological $21$~cm signal in collected data.  Terrestrially, for instance, one must contend with radio frequency interference from radio communications.  At these wavelengths, the ionosphere also interacts significantly with incoming photons, which may pose a challenge for instrumental calibration.  Perhaps most worrisome of all are astronomical foreground sources \citep{Oh, santos, Miguel, BowmanThesis,angelica,LOFAR}.  These act as contaminants to the signal and include unresolved extragalactic point sources, resolved bright point sources, and Galactic synchrotron radiation.  Taken together, these sources contribute a brightness temperature on the order of hundreds of Kelvins, which easily overwhelms the cosmological $21$~cm signals that are expected to be on the order of mK.  The ability to subtract off foregrounds is therefore crucial.

Foreground subtraction in 21-cm tomography is particularly challenging because current-generation instruments are not optimized for imaging.  Instead, most experiments are designed to provide cosmological information through \emph{statistical} measurements such as the power spectrum.  Images of the 21-cm sky are expected to be of a rather poor quality in terms of both signal-to-noise and synthesized beam, and image-processing techniques devised to optimally mitigate this problem may be too computationally intensive to be useful.  

In this paper, we take a conservative approach and investigate how well foreground cleaning can be done with a computationally 
simple method using only the the so-called \emph{dirty maps}  -- distorted maps of the sky as seen by an instrument at many different frequencies. Specifically, we clean out the foregrounds in any given sky direction by 
fitting a model of the contamination to the corresponding sky pixels in all the dirty maps.
This approach of cleaning one sky pixel at a time along the line-of-sight allows one to take advantage of the high spectral resolution offered by most 21~cm tomography experiments. The key idea is that explored in   \citet{xiaomin}: the point source emission (mainly synchrotron radiation is expected to vary smoothly with frequency
while the redshifted 21~cm signal of interest varies rapidly, because a small change in frequency (and hence redshift) corresponds to many Mpc in the radial direction and hence a major difference in local density fluctuations etc. High spectral resolution should thus allow one to separate out a rapidly oscillating cosmological signal from the spectrally smooth foreground using a low-order polynomial fit.

While previous studies \citep{xiaomin,LOFAR} have already highlighted the potential of this approach, such studies are incomplete because they do not take frequency dependent beam effects into account. 
At a given frequency, the dirty map is the true sky 
multiplied by a broad (say $10^\circ$) function known in radio astronomy as the 
{\it primary beam} and then convolved with a function referred to as the 
{\it synthesized beam}. 
The primary beam width is of order $\theta\sim\lambda/D_{\rm antenna}$, where $D_{\rm antenna}$ is the physical size of the individual array elements, whereas the synthesized beam is the Fourier transform of the 
array layout. Figure~\ref{awesome} shows a synthesized beam example illustrating several key features:
\begin{itemize}
\item There is a narrow central peak whose width is of order $\theta\sim\lambda/D_{\rm array}$, where $D_{\rm array}$ is the size of the whole antenna array. This peak width is usually referred to as the angular resolution of the experiment.
\item There is positive and negative ``frizz'' extending far beyond the central peak, corresponding to incomplete coverage of the Fourier plane, causing random-looking low amplitude oscillations on the angular resolution scale.
\item The synthesized profile is the same at all frequencies except for an overall scaling with the wavelength $\lambda$.
\end{itemize}
As one shifts to higher frequencies, the synthesized beam therefore shrinks like $\nu^{-1}$, causing the contributions from 
individual point sources to oscillate rapidly as positive and negative parts of the frizz moves across them.
This produces an effective foreground signal looking not like a smooth synchrotron spectrum, but more like the rapidly oscillating 
cosmological signal. 

Early work on this problem has provided cautionary but encouraging results \citep{BowmanThesis}.
Our goal in this paper is to quantify in detail how well the simple pixel-by-pixel foreground subtraction strategy can deal with this problem, using simulations including realistic radio arrays and beams.    \citet{Judd08} have independently studied this problem with a complementary approach, and we compare our results with theirs below. Whereas they perform a detailed case study of how well the MWA experiment can meet this challenge, including all foregrounds, we instead tackle the broader question of how residual point source contamination depends on experimental specifications, and what the experimental design implications are.  We ignore non-point-source contributions to the foregrounds, with the expectation that such components (such as Galactic synchrotron radiation) are rather benign relative to the unresolved sources.  Being spatially smooth, the off-beam contributions of such foregrounds tend to average out, unlike the point sources, even though their expected amplitude is larger.   \citet{Judd08} confirm this expectation.

In this paper, we deal only with the removal of \emph{unresolved} point sources, assuming that resolved and detected sources can
be eliminated by standard radio astronomy techniques as well as discarding the most contaminated pixels.
We thus envision the cleaning of foregrounds as a two-step process: first masking or deconvolving out bright resolvable point sources that exceed some flux cut, then proceeding with our proposed algorithm. We focus solely on the viability of the second step, but explore how the results depend on the flux cut from the first step.

The rest of this paper is organized as follows.  In Section \ref{method}, we outline the simulation methodology.  We go through the simulation of foregrounds, the simulation of radio interferometers, and the proposed foreground subtraction strategy itself.  The overall results of our analysis are presented in Section \ref{genresults}, where we consider three possible scenarios for foreground subtraction, ranging from the most pessimistic to the most optimistic. In Section \ref{howitdepends}, we quantify which design and algorithm choices are most important, and 
examine how the interplay between various instrumental and algorithmic parameters gives rise to these results. 
Table \ref{table1} lists the parameters that we explore as well as the ranges over which we vary them. 
In Section \ref{conc}, we summarize our results and discuss future prospects.

\begin{table*}
\label{table1}
\caption{Range of our parameter space exploration for foreground cleaning. 
Parameters pertaining both to experimental specifications and to analysis method impact the success of the cleaning.}
%\begin{ruledtabular}
\begin{tabular}{p{2.2cm} p{2.5cm}|p{3.0cm}|p{3.0cm}|p{3.0cm}}
\hline
\multicolumn{2}{c|}{\textbf{Assumptions}}  &  \textbf{Low-Performance Extreme} & \textbf{Fiducial Model} & \textbf{High-Performance Extreme} \\ \hline
\textbf{Experimental} & Tile Arrangement & $\rho(r) \sim r^{-2}$  & $\rho(r) \sim r^{-2}$  & Monolithic with tiles separated by $40\,\textrm{m}$\\ \cline{2-5}
& Rotation synthesis & None  & 6 hours, continuous & 6 hours, continuous \\ \cline{2-5} 
& Noise level & $\sigma_T \sim 1\,\textrm{mK}$  & Noiseless \footnotemark& Noiseless \\ 
\hline
\textbf{Analysis} & Primary beam width adjustments & None  & None  &  Adjusted to be frequency-independent\\ \cline{2-5}
& Bright point source flux cut $S_{\rm cut}$&100 mJy  &10 mJy &0.1 mJy\\ \cline{2-5}
& synthesized beam width adjustments & None & None  &Resolutions equalised by extra smoothing\\ \cline{2-5}
& $u$-$v$ plane weighting & None (natural)  & Uniform  &  Uniform \\ \cline{2-5}
& Order of polynomial fit & Constant  & Quadratic  &  Quintic \\ \cline{2-5}
& Range of polynomial fit & $80\,\textrm{MHz}$  & $2.4\,\textrm{MHz}$  &  $2.4\,\textrm{MHz}$ \\
\hline
\end{tabular}
%\end{ruledtabular}
\end{table*}

\section{Methodology}
\label{method}
Our basic approach is to simulate a point source sky, simulate observed maps of it at multiple frequencies, clean these maps and quantify the residual contamination that remains after the cleaning.
We repeat this a large number of times to explore the range design and algorithm options listed in Table~\ref{table1}.

A key feature of our cleaning method (which, again, is essentially that proposed in   \citet{xiaomin}, but with dirty map effects taken into account) is that the algorithm is blind, in the sense that our method does not rely on the any physical modeling of foregrounds.  This is important because the foregrounds relevant to $21$-cm tomography are still relatively poorly understood (as compared to, say, the foregrounds relevant to cosmic microwave background experiments), and models are often based on interpolations and extrapolations from other frequency bands \citep{angelica}.  Because of this, we choose to use only the generic property that radio frequency foregrounds are smooth functions of frequency.  This means that in our analysis, the foreground and instrumental \emph{simulation} steps are not only separate from each other, but also completely decoupled from the foreground \emph{subtraction} step.  Thus, in what follows we divide our description of the methodology into each of those steps.

\footnotetext{Our baseline simulations are run without noise because we show in Section \ref{genresults}
that the noise contribution can be included analytically and is unimportant for evaluating the effectiveness of our foreground subtraction strategies.}

\subsection{Step I: Simulation of Foregrounds}
While point sources tend to cluster and are therefore not randomly distributed, 
the clustering is rather weak, and for simplicity we make the the assumption that they are uncorrelated.
We thus simulate the contribution to each sky pixel independently, with its flux being the sum of the fluxes of a large number of of randomly generated point sources. 
The brightness of each point source is randomly drawn from the source count distribution
\begin{equation}
\frac{dn}{dS} = (4.0 \, \textrm{mJy}^{-1} \; \textrm{sr}^{-1} ) \left( \frac{S}{880 \, \textrm{mJy}} \right)^{-1.75},
\label{sourcecount}
\end{equation}
which is applicable at $\nu_{*} \equiv 150 \,\textrm{MHz}$ \citep{dimatteo1, lidz}.  The spectral dependence of each point source is given by
\begin{equation}
S(\nu) = S(\nu_{*} ) \left(\frac{\nu}{\nu_{*}} \right)^{-\alpha},
\end{equation}
where the spectral index $\alpha$ is randomly chosen from a Gaussian distribution 
\begin{equation}
p( \alpha ) = \frac{1}{\sqrt{ 2 \pi \sigma^2}} \exp \left[ - \frac{(\alpha - \alpha_0)^2 }{2 \sigma^2} \right],
\end{equation}
with $\alpha_0 = 2.5$ and $\sigma = 0.5$ \citep{max1}. Putting this all together, the brightness of each pixel is found by simply summing the simulated point sources within that pixel:
\begin{equation}
\label{Ieq}
I = \left( \frac{dB}{dT} \right)^{-1} \Omega_{sky}^{-1} \sum_{i=1}^{N} S_i^*  \left(\frac{\nu}{\nu_{*}} \right)^{-\alpha_i},
\end{equation}
where $S_i^{*}$ is the flux of the $i^{th}$ point source at $150\,\textrm{MHz}$, and $dB/dT = 6.9 \times 10^5 (\nu / \nu_{*})^2 \, \textrm{mJy} \, \textrm{K}^{-1} \, \textrm{sr}^{-1}$ is the standard conversion between intensity and brightness temperature. We use a pixel size of $\Omega_{sky} = 4.3 \, \textrm{arcmin}^2$ in our simulations.  In generating our point sources, equation \ref{sourcecount} is truncated at some maximum flux $S_{max}=S_{cut}$ because the brightest point sources are assumed to be detected and separately removed as mentioned above.  We leave the threshold $S_{cut}$ as a free parameter, and we investigate how the effectiveness of our foreground subtraction algorithm changes as $S_{cut}$ is varied from $0.1\,\textrm{mJy}$ to $100\,\textrm{mJy}$.  At the dim end of the distribution, we truncate at a minimum flux $S_{\rm min}$ because the source count distribution diverges as $S \rightarrow 0$.  We find that it is unnecessary to have $S_{\rm min}$ be any lower than $10^{-3} \,\textrm{mJy}$, as the total flux $I$ has converged by then.  These values for $S_{\rm min}$ and $S_{max}$ mean that we typically simulate $N \sim 2000$ point sources point sources for the sum in equation \ref{Ieq}.  

\begin{figure}
\centering
\includegraphics[width=0.5\textwidth]{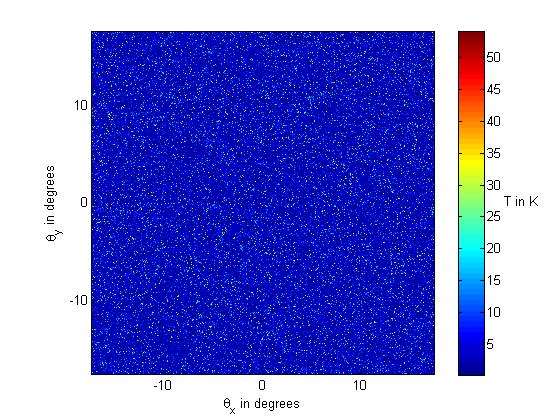}
   \caption{A sample sky with point source foregrounds at $\nu = 159\,\textrm{MHz}$}
   \label{samplesky}
\end{figure}

All the simulations were performed on $1024 \times 1024$ grids, and we verified that increasing the resolution to $4096 \times 4096$ had essentially no effect on the results.  In the line-of-sight direction, the lowest frequency that we simulate is $\nu = 158.73\,\textrm{MHz}$ (corresponding to $z=8$), and (except for in section \ref{rangeofpolyfit}) the frequency increases by $\Delta \nu = 0.03 \,\textrm{MHz}$ from one frequency slice to another over $80$ slices.  In section \ref{rangeofpolyfit} we continue to simulate $80$ slices starting at $\nu=158.73\,\textrm{MHz}$, but we increase $\Delta \nu$ to $0.3\,\textrm{MHz}$ and $1.0\,\textrm{MHz}$ to quantify the effect of increasing the range of the polynomial fit.  A sample sky at $\nu = 159\,\textrm{MHz}$ is shown in figure \ref{samplesky}.

Performing the sum in equation \ref{Ieq} at all frequencies for all $1024^2$ pixels would take on order a month with our software.  We therefore accelerate our algorithm by exploiting the fact that the pixels are independent random variables. 
We first generate a database of 1000 pixels for which we compute the full frequency dependence. 
We then give each sky pixel the frequency dependence of a random pixel from the archive. It is easy to show that this procedure does
not bias the residual power spectra low or high, but merely increases the variance in our estimation of these power spectra.  By computing the scatter between multiple analyses using different random seeds and database sizes, we find that this excess scatter becomes unimportant for our purposes once the database exceeds a few hundred pixels.

We re-emphasize that the process used here pertains only to the \emph{simulation} of foregrounds.  The foreground subtraction itself is blind and therefore does not depend on the assumed properties of the foregrounds.
\subsection{Step II: Simulation of Radio Interferometer}
\label{simstep}
As mentioned above, the response of a radio interferometer array can be described by two beam functions: the primary beam and the synthesized beam. The primary beam function encodes the power response of an individual interferometer element to different parts of the sky, while the synthesized beam is a convolution kernel that describes the interferometric effects of the array as a whole.  To a good approximation, what an interferometer sees is not the true sky, but the true sky multiplied by the primary beam and then convolved with the synthesized beam.

In our simulations, we approximate the primary beam by a Gaussian.
Its width scales as $\lambda/D_{\rm antenna}$, and is chosen so that the width matches simulated antenna patterns of the Murchison Widefield Array (MWA) dipole antennas at $150$ MHz (approximately $30^{\circ}$ full-width-half-max).
The synthesized beam is found by taking the Fourier transform of the distribution of baselines
(in the so-called $u$-$v$ plane), which depends on the arrangement of tiles in the interferometer array. While our simulations of course require specific realizations of the array layout, we expect our results to hold for any radio array whose tile distribution is qualitatively similar to the various scenarios we consider in the results section.  Roughly speaking, this means that our results should be applicable to any array with a relatively large number of short baselines and a number of long baselines that span up to $\sim 1\,\textrm{km}$, giving an angular resolution that is on the order of arcminutes.
For an array with very different dimensions, our results can be straightforwardly scaled.

Integration time affects the response of the interferometer array in two ways that are important here. First, detector noise averages down with time as $t^{-1/2}$ as long as it is uncorrelated over different time periods.
A typical point source sensitivity for a current-generation experiment like the MWA is $S= 0.27\,\textrm{mJy}$, with a $4.6$ square arcminute pixel, and a $32\,\textrm{MHz}$ bandwidth.  This corresponds to an r.m.s. detector noise per a pixel of
\citep{xiaomin}:
\begin{equation}
\sigma_T^{\textrm{MWA}} = 0.22\,\textrm{K} \left( \frac{32\textrm{MHz}}{\Delta \nu} \right)^{1/2} \left( \frac{1\,\textrm{hour}}{t} \right)^{1/2},
\end{equation}
where $\Delta \nu$ is the bandwidth, and $t$ is the integration time.  
The MWA has a channel bandwidth of $32\,\textrm{kHz}$, 
and with $\sim\!\!\!1000$ hours of integration the detector noise level is $\sim 0.2\,\textrm{K}$.  In our simulations we consider not only current-generation experiments with this level of noise, but also the noise levels of hypothetical future experiments, which we take to be of order $\sim 1\,\textrm{mK}$.

\begin{figure}
\centering
\includegraphics[width=0.5\textwidth]{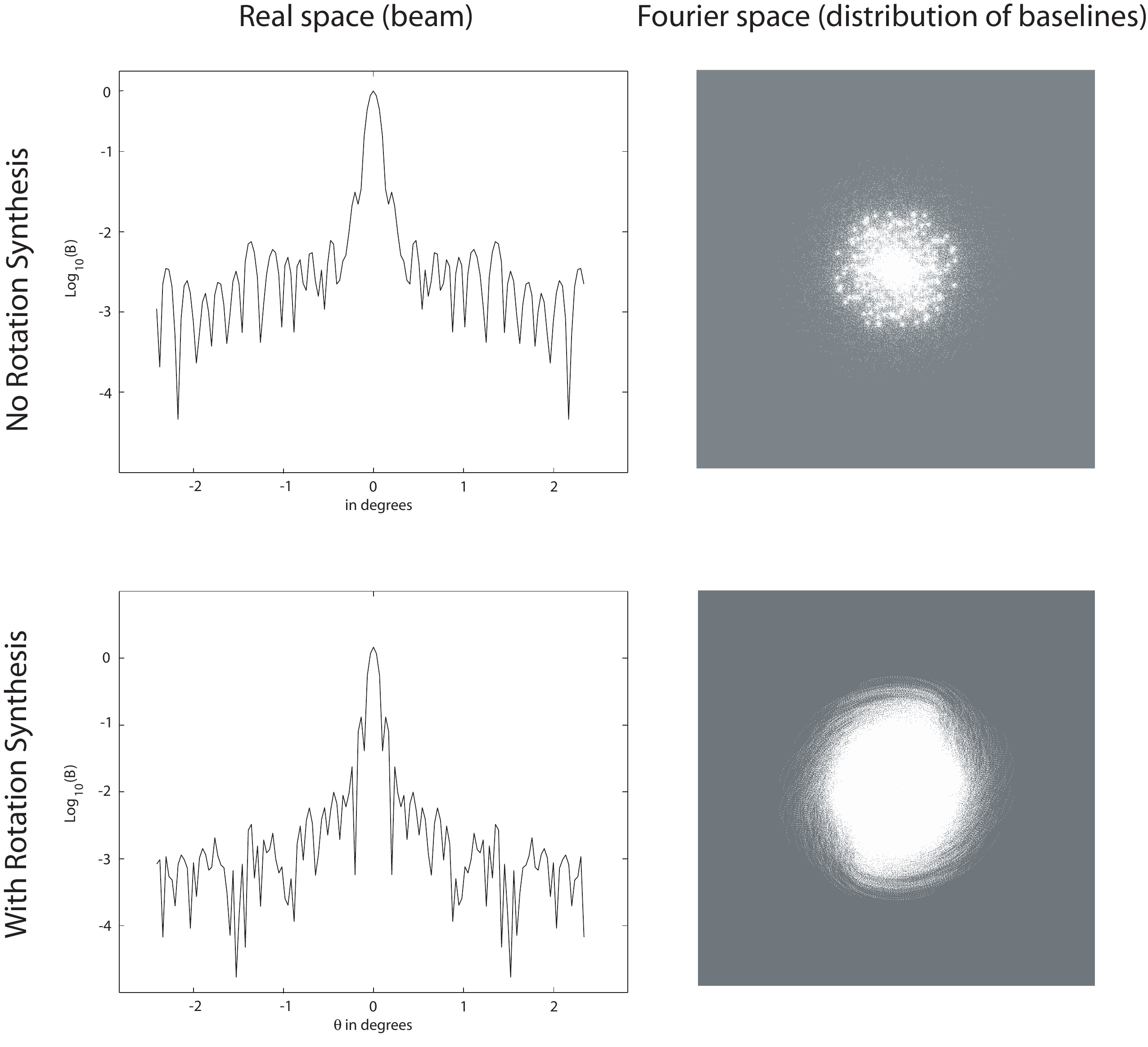}
   \caption{The left hand column shows sample beam profiles (a real-space description of the beam) while the right hand column shows the corresponding $u$-$v$ distribution of baselines (a Fourier-space description of the beam). The top row illustrates an array with no rotation synthesis, while the bottom row shows an array with $6$ hours of rotation synthesis.  The real-space beams are normalized so that their peaks are at $1$.}
\label{pedfourier}
\end{figure}

The second effect of taking measurements over an extended period of time pertains to the concept of rotation synthesis.  The rotation of the Earth during observations means that interferometer baselines are not static points on the $u$-$v$ plane, but instead sweep out of $u$-$v$ tracks.  This results in a change in the properties of the synthesized beam, as one can see in figure \ref{pedfourier}.  In our analysis, we explore a range of rotation synthesis scenarios from none at all (i.e. taking a snapshot of the sky) to having $6$ hours of rotation synthesis.\\

\subsection{Step III: Foreground Subtraction}
As mentioned above, we use a pixel-by-pixel line-of-sight cleaning strategy for removing the unresolved foregrounds below the flux threshold.  For each pixel, its frequency dependence is fit by a 
low-order polynomial.  Such a polynomial is by definition a smooth function, and thus the hope is that by subtracting off the fit from the signal, one subtracts off mainly the smooth foregrounds and not much of the cosmological signal, which is expected to oscillate wildly with frequency. We leave the order of the polynomial as a free parameter, and explore the effects of varying it from 0 (constant) to 5 (quintic).
This involves a tradeoff between two separate effects.  On one hand, the higher the order of the fit, the lower the residual foregrounds will be. On the other hand, high order fits come at the expense of fitting more power out of the cosmological signal. Crudely speaking, one therefore wishes to select the \emph{lowest} order capable of adequately subtracting off foregrounds.

\section{Results}
\begin{figure}
\centering
\includegraphics[width=0.5\textwidth]{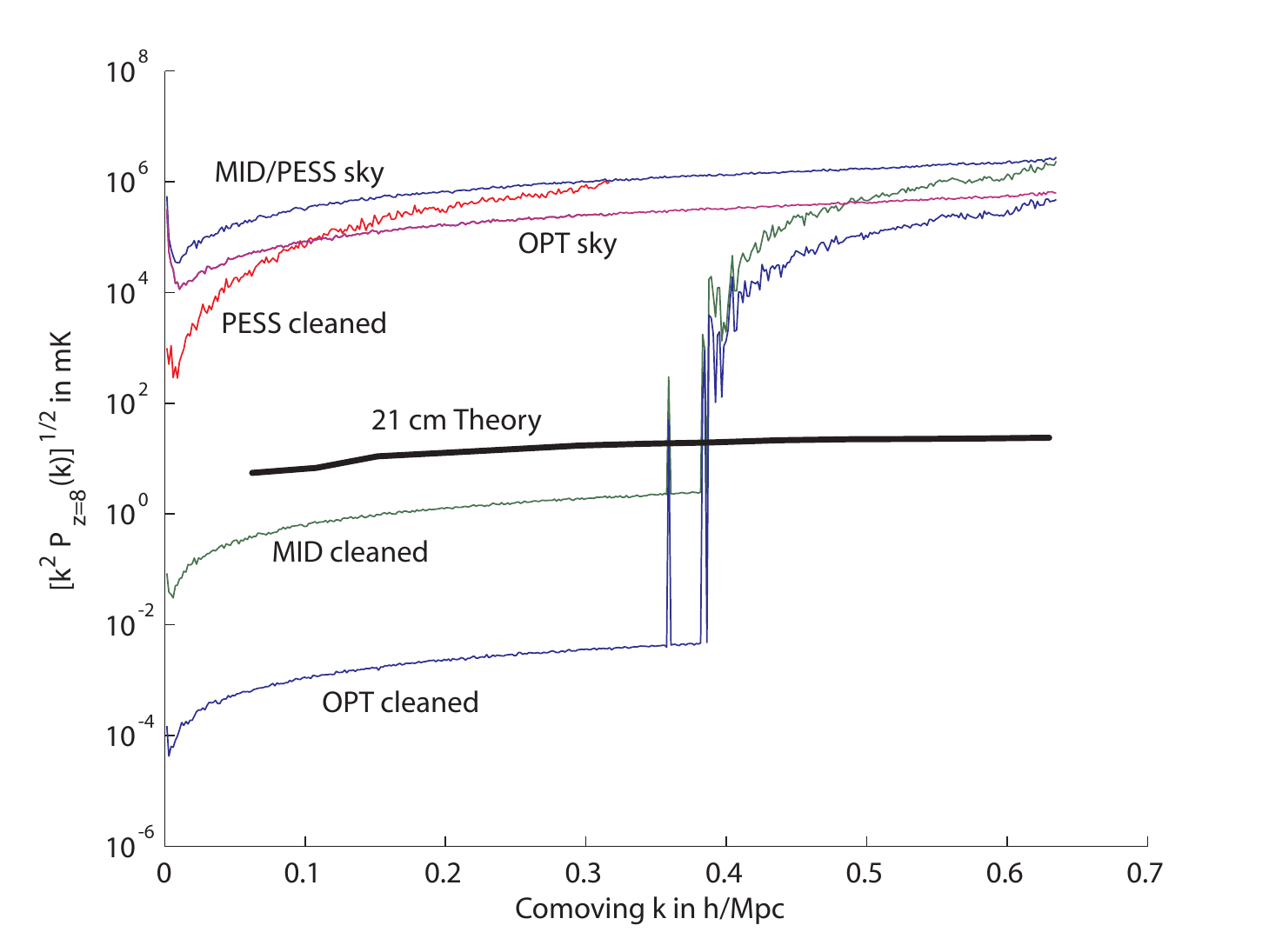}
\caption{2D power spectra of foregrounds and foreground residuals for the various scenarios outlined in the text.  It is clear that except for the most pessimistic scenario, a dirty-map pixel-by-pixel cleaning strategy can be used to get within at least striking distance of the cosmological signal.}
\label{summary}
\end{figure}
In this section, we quantify the extent to which point source contamination can be removed with line-of-sight cleaning, and how this depends on the many parameters in Table~1.  To get a sense of the extent to which the parameters matter, we then analyze three scenarios designed to bracket the range of possibilities in Section~\ref{ThreeScenarios}. As seen in Figure~\ref{summary}, this range is broad, extending all the way from utter failure to recover the cosmological signal to success in pushing foregrounds two orders of magnitude below the signal.  After discussing some general findings that are independent of all these parameters in Section~\ref{genresults}, we devote Section~\ref{howitdepends} to examining the parameters one by one to quantify which ones make the greatest difference.

\subsection{Three scenarios}
\label{ThreeScenarios}

We now analyze three scenarios designed to bracket the range of possibilities:
\begin{enumerate}
\item \textbf{Pessimistic scenario (PESS)}
\begin{itemize}
\item \emph{Experimental assumptions:} In this scenario, we simulate an array whose tiles are arranged in a radial density profile that goes as $r^{-2}$.  The array takes a single snapshot of the sky.
\item \emph{Analysis:} No extra weighting is imposed on the $u$-$v$ plane (see Section \ref{weights} for details on weighting schemes), and no attempt is made to adjust for the fact that both the primary beam and the synthesized beam have frequency-dependent widths.  It is assumed that only sources that are $10\,\textrm{mJy}$ or brighter can be removed prior to the subtraction of unresolved point sources.  We also assume that the highest order polynomial that one can fit without taking out significant power from the cosmic signal is a quadratic.
\end{itemize}
\item \textbf{Fiducial scenario (MID)}  --- this scenario is designed to be reasonably representative of current-generation experiments.
\begin{itemize}
\item \emph{Experimental assumptions:} Same as PESS except that the array samples the $u$-$v$ plane continuously for $6$ hours.
\item \emph{Analysis:} Same as PESS except that the $u$-$v$ plane measurements are weighted so that all parts of the $u$-$v$ plane that are covered are given a uniform weight.
\end{itemize}
\item \textbf{Optimistic scenario (OPT)}
\begin{itemize}
\item \emph{Experimental assumptions:} Same as MID.
\item \emph{Analysis:} Same as MID except that we assume that bright point sources can be removed down to $1.0\,\textrm{mJy}$, and that one can safely fit up to a cubic polynomial.
\end{itemize}
\end{enumerate}
Our fiducial model is a middle-of-the-road (MID) case intended to be representative of current generation experiments, while the pessimistic (PESS) and optimistic (OPT) scenarios refer to worst and best case experimental expectations, respectively. 
The PESS and OPT scenarios in general differ less than the low-performance and high-performance extremes outlined in Table~\ref{table1} because the parameters of these extreme scenarios are in some cases unrealistic, and are considered in 
Section~\ref{howitdepends} solely to better understand how the various parameters affect foreground subtraction.  

As our measure of how well or poorly the cleaning works, we use the two-dimensional spatial power spectrum of the cleaned map. We do this because the power spectrum of the cosmological signal is the key quantity that the 21 cm tomography community aims to measure in the near term, and the residual point source power spectrum will simply add to this.  In all our power spectrum plots (starting with Figure~\ref{summary}), we have removed the distorting effect of convolution with the synthesized beam, so that $P_{2D}(k)$ for the point sources is a constant ($k$-independent) white noise spectrum.  The detector noise also has a constant spectrum, and the cosmological 21 cm signal plotted is what would be seen by an ideal instrument making a distortion-free image of the sky. The deconvolution step is trivial to perform in Fourier space, corresponding simply to dividing the measured power spectrum by the radial distribution of baselines.

Rather than plot $P_{2D}(k)$ itself, we plot the dimensionless quantity $[k^2 P_{2D}(k)]^{1/2}$, which can be roughly interpreted as the dimensionless fluctuation lever, analogous to the standard quantity 
$\delta_T/T\propto [\ell^2 C_\ell]^2$ for the cosmic microwave background and the quantity
$\Delta\propto [k^3 P_{3D}(k)]^{1/2}$ for three-dimensional galaxy surveys.  In the figures that follow, $[k^2 P_{2D}(k)]^{1/2}$ is always plotted for the frequency slice at $\nu=158.73\,\textrm{MHz}$. This translates to a redshift $z=8$, around the ``sweet spot" of most current generation 21 cm experiments.

Our cleaning method is linear, i.e., the ``data cube'' containing the cleaned maps at different frequencies is simply some linear combination 
of the input data cube, and the weights in this linear combination (implementing the polynomial fits) 
are fixed, independent of what the data cube contains.
This means that if we have three data cubes, containing cosmological 21 cm signal, detector noise and point source signal, respectively, we
get the same result if we sum them and then clean as if we first clean them individually and then sum them.
Since these three components are all statistically independent, this also implies that their post-cleaning power spectra simply add.
We take advantage of this fact (which we also verify numerically 
in Section~\ref{noise}) to simplify our analysis, performing most of our simulations with both the cosmological signal and the noise set to zero. A perfect foreground subtraction algorithm would thus produce a residual power spectrum that is zero everywhere.

Figure~\ref{summary} shows the results.
For comparison, the figure also shows the simulated EOR signal for $z=8$ (taken from \citet{Matt2} and \citet{Matt1}). 
Figure \ref{summary} shows that except for in the PESS scenario, our simple foreground subtraction algorithm is successful on large scales but not on small scales. Let us now focus on understanding this better, and clarifying the key properties of the cleaning method.

\begin{figure}
\centering
\includegraphics[width=0.5\textwidth]{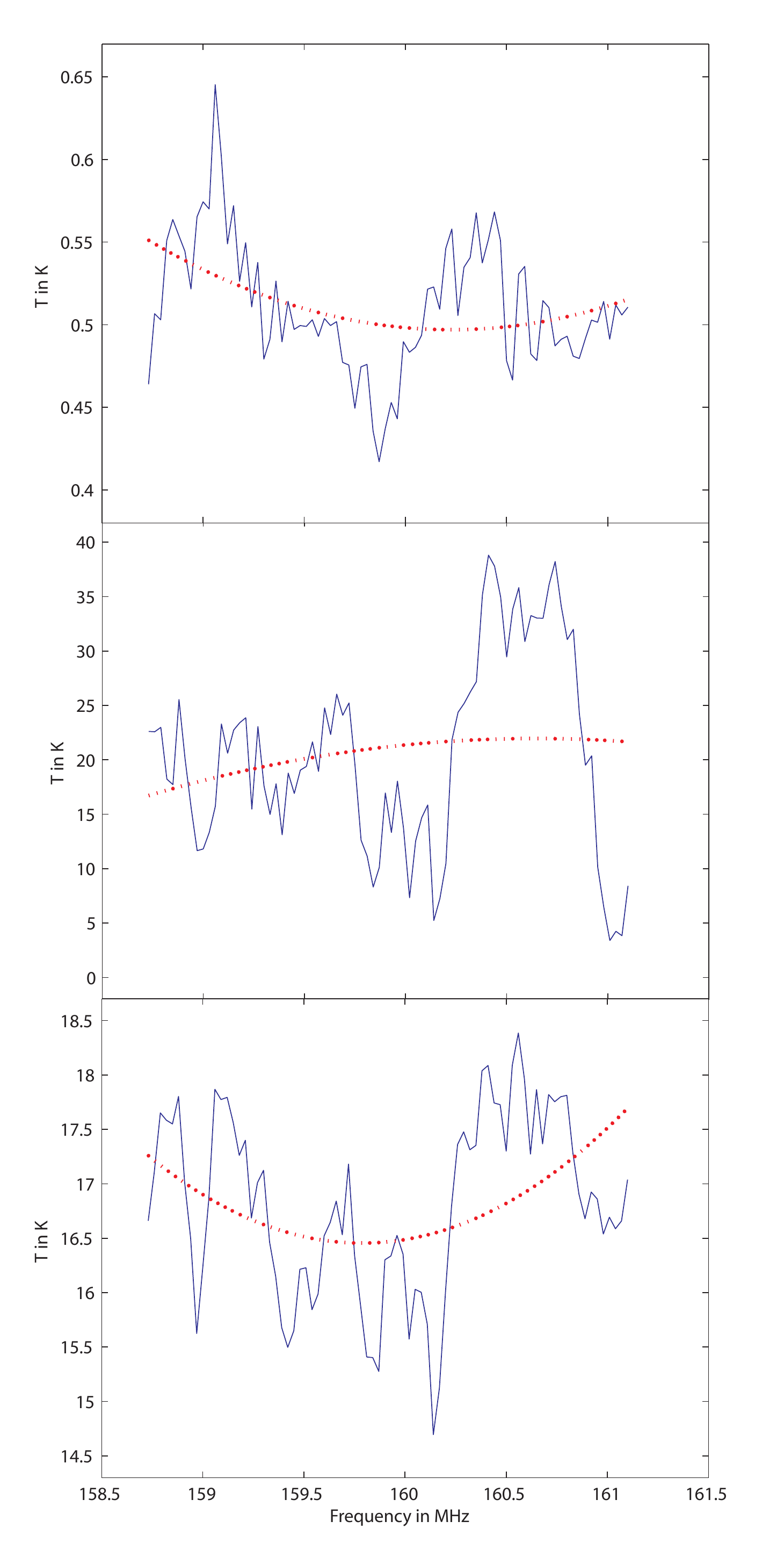}
\caption{The spectra of a three typical dirty-map pixels, with fits given by the dotted curves.  The top panel assumes no instrumental noise and no point sources brighter than $0.1\,\textrm{mJy}$.  The middle panel assumes no instrumental noise and no point sources brighter than $100.0\,\textrm{mJy}$.  The bottom panel assumes an instrumental noise level of $\sigma_T = 200 \,\textrm{mK}$ and no point sources brighter than $0.1\,\textrm{mJy}$.}
\label{combinedspec}
\end{figure}

\subsection{Why the method can work well on large scales}
\label{genresults}

In the top panel of figure \ref{combinedspec}, we show the spectrum for a pixel randomly chosen from the dirty map produced by a noiseless but otherwise MWA-like instrument that has been continuously observing a fixed patch in the sky for $6$ hours.  A bright point source flux cut of $S_{cut}=0.1\,\textrm{mJy}$ has been assumed.  In figure \ref{combinedspec} we can see that while the overall trends of the spectrum can be captured by a low-order polynomial fit, the finer features generally remain as residuals.

The effectiveness of the polynomial fit depends on a variety of factors.  We find that the greatest variation comes from varying $S_{cut}$ and the noise level.  In the middle panel of figure \ref{combinedspec}, we show a typical pixel with $S_{cut}=100.0\,\textrm{mJy}$\footnote{Note that in general, dirty-map pixels may be negative.  Such negative brightness temperatures arise from the fact that in radio arrays, the signal from a single tile is never correlated with itself.  We thus never sample the origin of the $u$-$v$ plane, and therefore lose the mean of the signal.}.  Similarly, the addition of noise can decrease the quality of the fit. In bottom panel of figure \ref{combinedspec}, we have added $\sigma_T = 200\,\textrm{mK}$ (note the larger amplitudes on the $y$-axis compared to what is shown in the top panel).

From figure \ref{combinedspec}, the reader may be surprised that our subtraction strategy works at all. Indeed, it seems from the figure that low order polynomial fits do extremely poorly, and that most of the foregrounds will remain after the subtraction. 
So why, then, does Figure~\ref{summary} show that the foregrounds can be suppressed by 
many orders of magnitude?

The above-mentioned linearity holds the key to the success. 
To understand why, consider again the 3-dimensional data cube, and imagine it arranged so that 
the 2-dimensional dirty maps are all horizontal, stacked vertically on top of one another so that
the vertical direction corresponds to frequency.
To generate a plot such as Figure~\ref{summary}, we perform two operations on this cube of simulated data:
\begin{enumerate}
\item Clean one pixel (vertical column) at a time 
\item Fourier transform one frequency map (horizontal slice) at a time
\end{enumerate}
Since both of these steps are linear, they can each be thought of as multiplying all the data (rearranged into a single vector) by some matrix. Since the two steps mix data purely horizontally and purely vertically, respectively, it is easy to see that the corresponding two matrices must commute. In other words, we get the same result if we perform the steps in the opposite order (our simulations confirm this).
An analogous and more familiar example is the 3-dimensional Fourier transform, which decomposes into successive Fourier transforms 
in the vertical and two horizontal directions, and again gives the same result regardless of the order in which these operations are performed.

\begin{figure}
\centering
\includegraphics[width=0.5\textwidth]{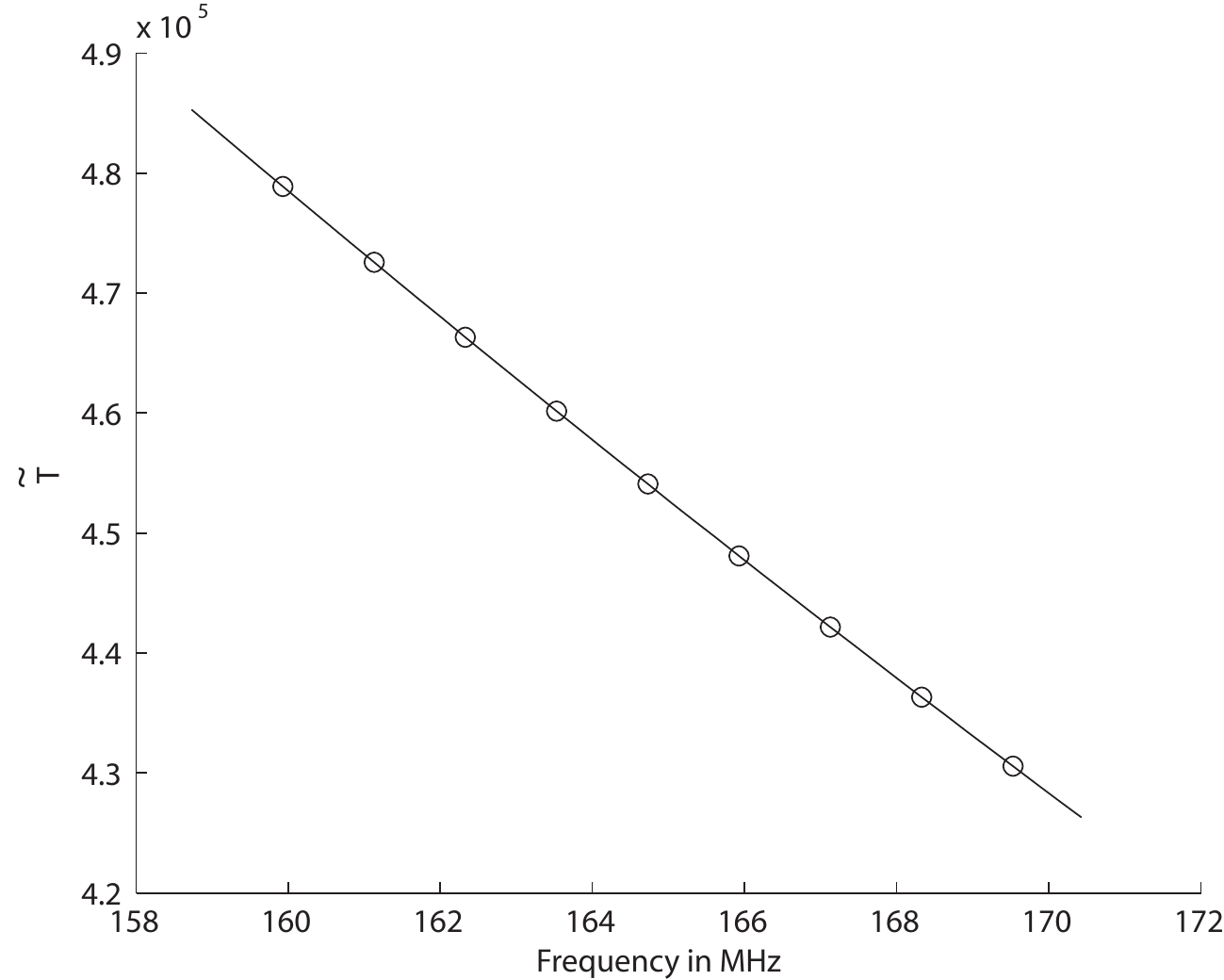}
\caption{The spectrum of a Fourier space pixel, taken from the part of the plane where $u$-$v$ coverage is complete.  A fit to the spectrum is shown.}
\label{combined}
\end{figure}

This means that we get the exact same result in Figure~\ref{summary} if we first Fourier transform the maps, then perform the cleaning 
one Fourier mode at a time instead of one pixel at a time.
Figure \ref{combined} shows the spectral fit to a sample pixel in on the Fourier ($u$-$v$) plane, revealing a very smooth curve that
can be fit with exquisite accuracy. The success for this particular Fourier mode hinges on the fact that it lies in the central part of the Fourier plane which has been well sampled by the array at all frequencies. This explains the low plateau in the residual power spectra in the left side of Figure~\ref{summary}. The sharp rise in residual power on smaller scales corresponds to incomplete 
Fourier coverage, with certain interferometric baselines missing.  Converting this intuitive understanding back to real space, the small-scale synthesized beam frizz seen in Figure~\ref{awesome}
is largely averaged away when expanding the sky into long-wavelength Fourier modes, whereas the small-scale modes are severely affected.  The characteristic scale separating ``short" and ``long" Fourier modes is determined by the \emph{longest baseline radius for which $u$-$v$ coverage is complete}.  As one moves outwards from the origin in Fourier space, one eventually reaches a pixel where $u$-$v$ coverage is incomplete at some frequency along the line of sight.  Once there is any such missing information along the frequency direction, our simple fitting algorithms fail to find a good fit and the result is a large increase in residuals.  Since this increase is caused by incomplete $u$-$v$ coverage, the scale at which this occurs depends on the rotation synthesis scenario being considered.  We find, however, that for most reasonable cases the upswing takes place between $k=0.3h/\textrm{Mpc and }0.4h/\textrm{Mpc}$.

The reader may also be concerned (based on figure \ref{combinedspec}) that there may be much residual power in the frequency direction, leading to problems for experiments aiming to measure the 3D power spectrum.  A full discussion of the 3D power spectrum is beyond the scope of this paper, but for now we note that plots like figure \ref{combinedspec} are simply \emph{real space} plots in the line-of-sight direction.  
Fourier transforming in the perpendicular direction, the fit becomes quite good in the central parts of the
$u-v$ plane as mentioned above.
Fourier transforming in all three directions (which is what one would ultimately do in order to find estimate the 3D power spectrum), one does not expect the residual power to 
depend on both the line-of-sight and transverse wave number. 
Indeed, \citet{Judd08} find that while there is some contamination along the line-of-sight, there do exist clean regions in the full 3D Fourier space.

Aside from residual foregrounds, another problem that may arise is the fact that blind algorithms such as ours have no way to distinguish between signal and foreground except for through differences in smoothness as a function of frequency.  Thus, if the signal possesses any smooth large-scale component, the foreground subtraction algorithm may inadvertently remove this component of the signal with the foregrounds.  Deciding whether this is a problem or not (and quantifying the seriousness if it \emph{is} indeed a problem) requires the detailed examination of the properties of the signal, which is beyond the scope of this paper.  Intuitively, however, one can say (at the very least) that there will be no loss of cosmological signal in the transverse Fourier components, since our algorithm subtracts along the line-of-sight.

\subsection{Exploration of parameter space}
\label{howitdepends}

Above we saw that there was a huge difference in foreground removal ability between the three scenarios. 
Let us now investigate which of the instrumental and algorithmic parameters in Table~\ref{table1} make the greatest difference. 
We do this by varying one of them at a time, over the range given in Table~\ref{table1}, while keeping all others fixed.

\begin{figure}
\centering
\includegraphics[width=0.5\textwidth]{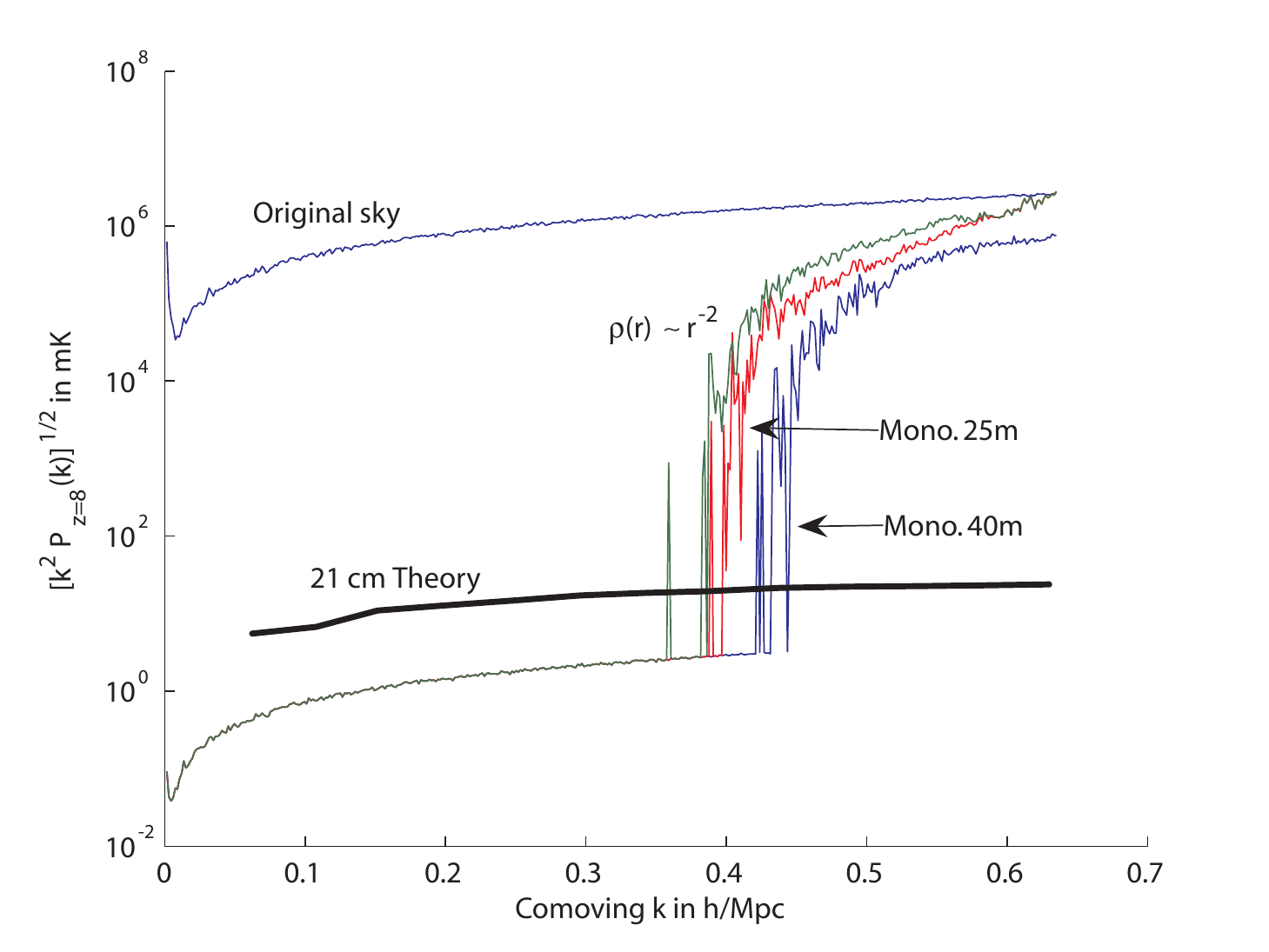}
\caption{Dependence on array layout: 2D power spectra for the fiducial model with various arrangements of tiles. Monolithic and $r^{-2}$ arrangements both seem to work well.}
\label{tilearrangement}
\end{figure}

\subsubsection{Array Layout}

In a radio array, it is the arrangement of antenna elements that determines the distributions of baselines, and it is the distribution of baselines that determines the shape of the synthesized beam.  Thus, our ability to subtract foregrounds depends strongly on the layout of the array.

In an experiment such as the MWA, each antenna element consists of a tile of 16 crossed dipole antennas, arranged in a 4 by 4 grid with the dipoles spaced roughly $1\,\textrm{m}$ from each other.  In this configuration, the dipoles give rise to a complicated primary beam pattern which may include structures such as sidelobes.  As mentioned in previous sections, we neglect these complications and assume that each tile of dipoles has a primary beam that takes the form of a Gaussian with a $30^{\circ}$ full-width-half-max.  

We consider two different types of tile arrangement.  One has the density of tiles varying as $r^{-2}$, where $r$ is the radius from the centre of the array.  The other, which we categorize as a ``monolithic" arrangement, is a regular square grid densely packed near the centre of the array.  
A few tiles are placed farther from the centre to provide some long baselines for calibration purposes and for good angular resolution.  In both cases, we have a total of $500$ tiles in our simulations; for the monolithic cases $400$ tiles form the central core, while $100$ tiles scattered outside the core provide some short baselines (which can be formed between two close-by tiles that are both outside the core) and many long baselines (which are formed by pairing up a tile within the core to one outside the core).  We consider two realizations of the monolithic cases, one with tiles in the central core separated by $25\,\textrm{m}$ and with them separated by $40\,\textrm{m}$.  In both cases, the combination of short-to-medium-length baselines formed between tiles in the core and the extra few short baselines provided by closely separated tiles outside the core allow complete coverage of the $u$-$v$ plane near the origin with the help of rotation synthesis.

The results are shown in figure \ref{tilearrangement}.  From the plot, it is clear that good foreground subtraction can be done using either an $r^{-2}$ arrangement or a monolithic arrangement, and that on large scales there is no difference in performance.  Monolithic arrangements, however, seem capable of pushing foreground subtraction to slightly finer scales.  The explanation for this is that with an $r^{-2}$ arrangement, the high density of tiles at small $r$ means that the inner parts of the $u$-$v$ plane are vastly oversampled given the rotation synthesis, and one can spread out the tiles slightly to extend the perfectly covered regions of the $u$-$v$ plane without introducing any gaps thanks to the tight monolithic core.

\begin{figure}
\centering
\includegraphics[width=0.5\textwidth]{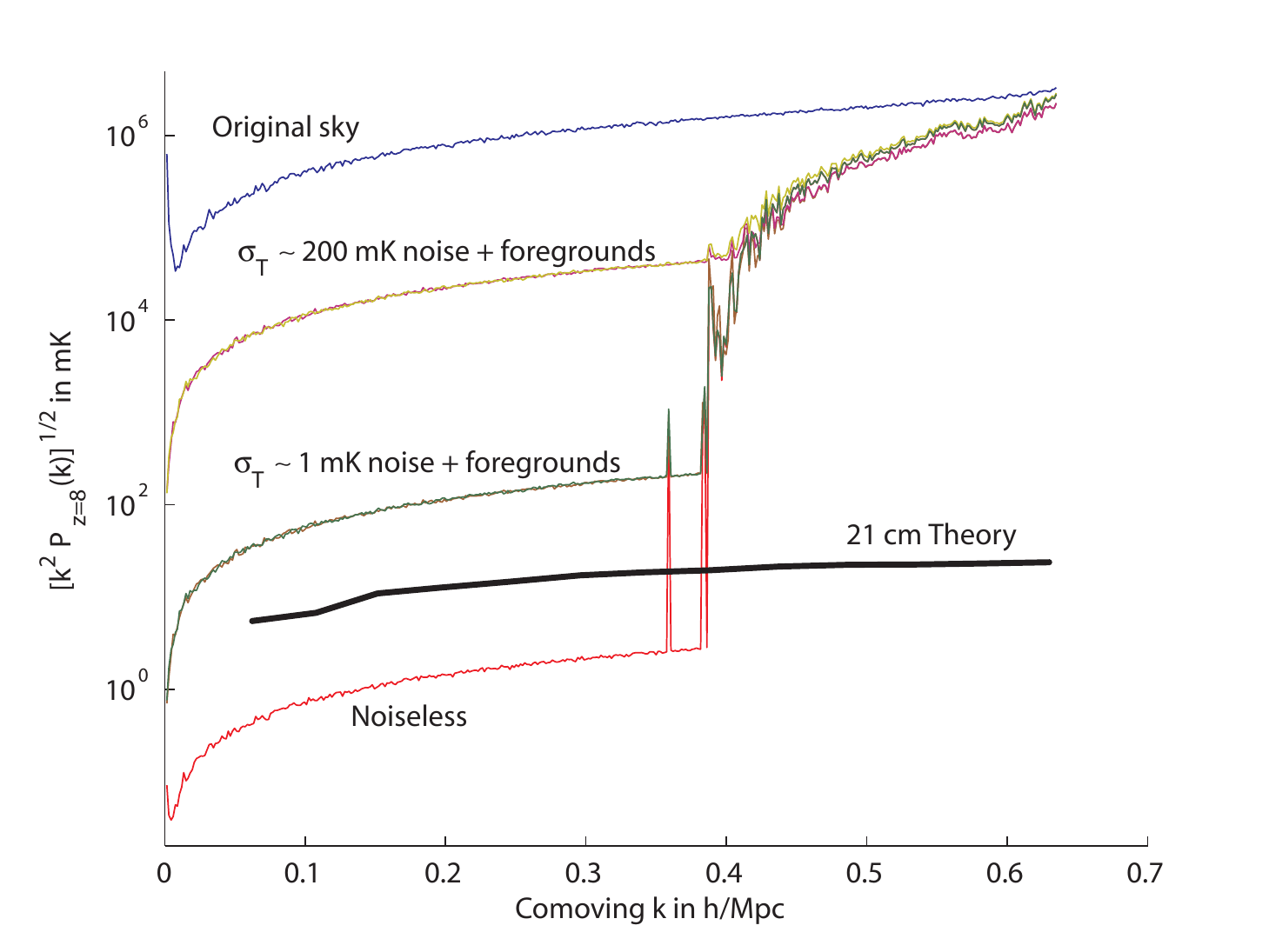}
\caption{Effect of noise: 2D power spectra for the fiducial model with various levels of instrumental noise show that the power spectrum of detector noise simply gets added to the power spectrum of the residual foregrounds.}
\label{noisefig}
\end{figure}

\subsubsection{Noise Level}
\label{noise}

As mentioned above, we consider three different scenarios in our analysis of noise effects: a noise level $\sigma_T \sim 200\,\textrm{mK}$ (representative of current-generation instruments), a noise level $\sigma_T \sim 1\,\textrm{mK}$ (hopefully representative of next-generation instruments), and a hypothetical noiseless case.  The resulting power spectra are shown in figure \ref{noisefig}.
For all but the noiseless case, there are in fact two sets of curves plotted:
one showing the result of the simulation with noise, and the other showing the noiseless simulation with the input noise power spectrum (a constant) 
added afterward.
These two sets of curves lie on top of each other and are visually indistinguishable, confirming that our linear cleaning method leaves the noise power 
essentially unaffected. It is therefore unnecessary to run simulations with noise, since the results can be predicted analytically given a noiseless simulation.

A naive reading of Figure~\ref{noisefig} suggests the pessimistic conclusion that current-generation 21~cm tomography experiments have no hope of seeing a cosmological signal. However, the additive contribution to the power spectrum from noise can be completely eliminated by measuring the power spectrum by cross correlating maps made at different times, since their noise will be uncorrelated. The WMAP team have successfully used an analogous procedure for noise bias removal, where they cross correlated maps made not at different times but with different receivers \citep{cmb}. In contrast, the contribution from residual point sources can not be eliminated in this way.

Just like for WMAP, the noise will still contribute to the {\it error bars} 
$\Delta P(k)$ on the measured power spectrum, but these error bars can be shrunk by averaging many Fourier modes with comparable wave number $k$. For the case where noise and signal have a Gaussian distribution, 
$\Delta P(k)=\sqrt{2/N} P(k)$ where $P(k)$ is the total power spectrum (including noise and residual point sources)
and $N$ is the number of Fourier modes averaged. For example, the noise power exceeds the signal power around 
the third acoustic peak of the CMB power spectrum measured by WMAP, but $N$ is large enough that the error bars nonetheless become significantly smaller than the cosmic signal.  In our case, binning radially in annuli of thickness $k=0.1h/\textrm{Mpc}$, we typically have $N \sim 10^4\textrm{ to } 10^5$ (depending the value of $k$), which would suggest percent level relative errors on the plotted power spectrum curves.

In summary, our results regarding noise are quite encouraging: the promising forecasts that have been made in the past for what can be learned from 21 cm cosmology all included detector noise, but not the point source contribution in the full complexity that we are modeling. Our results show that the fact the point sources can be removed without
having much effect on the noise levels. 

\begin{figure}
\centering
\includegraphics[width=0.5\textwidth]{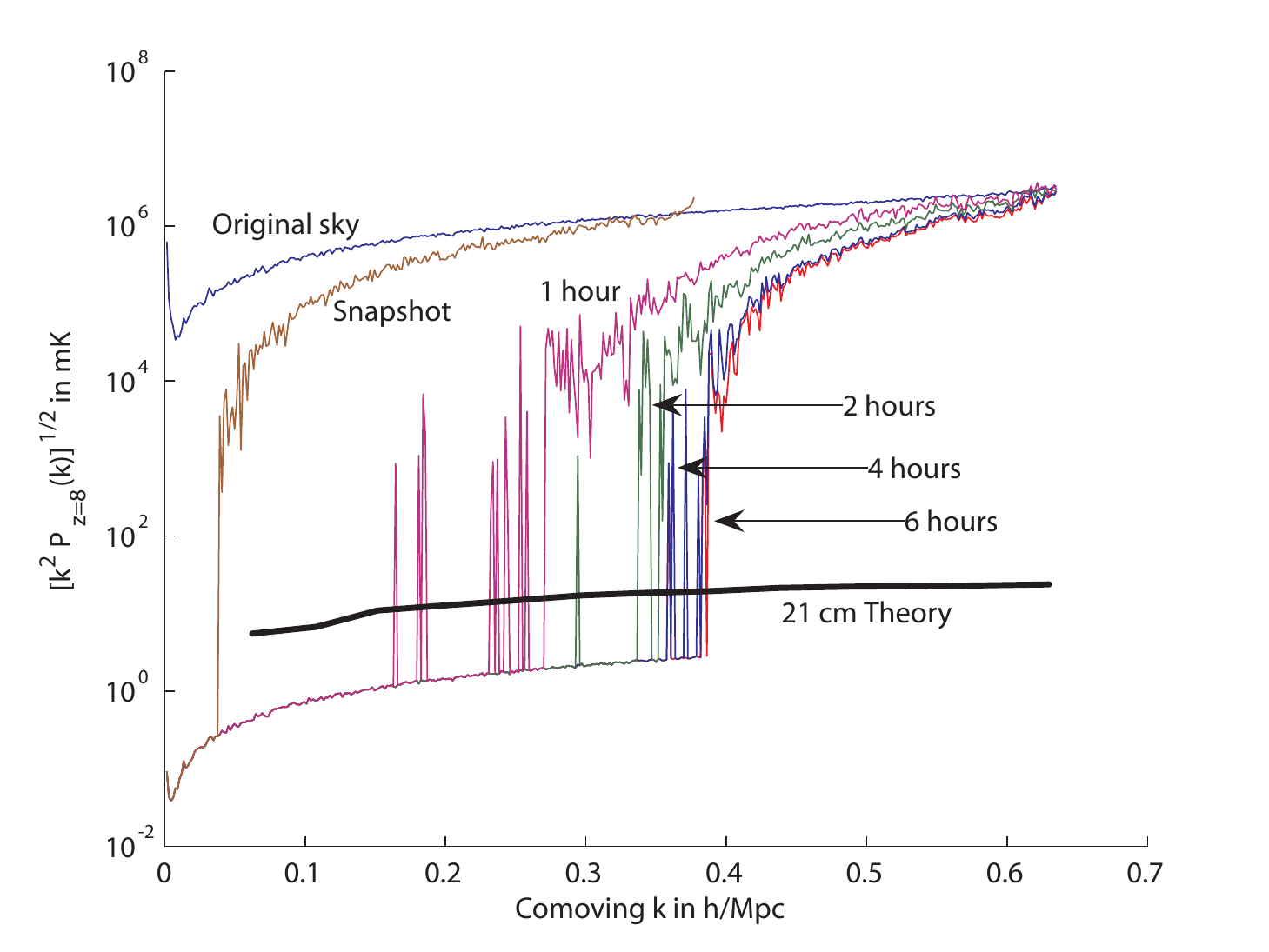}
\caption{Effect of rotation synthesis: 2D power spectra are show for the fiducial model but with various total rotation synthesis times. The effect of lengthening integration time to increase $u$-$v$ coverage is seen to saturate after about 4 hours.}
\label{totalrotsynth}
\end{figure}

\begin{figure}
\centering
\includegraphics[width=0.5\textwidth]{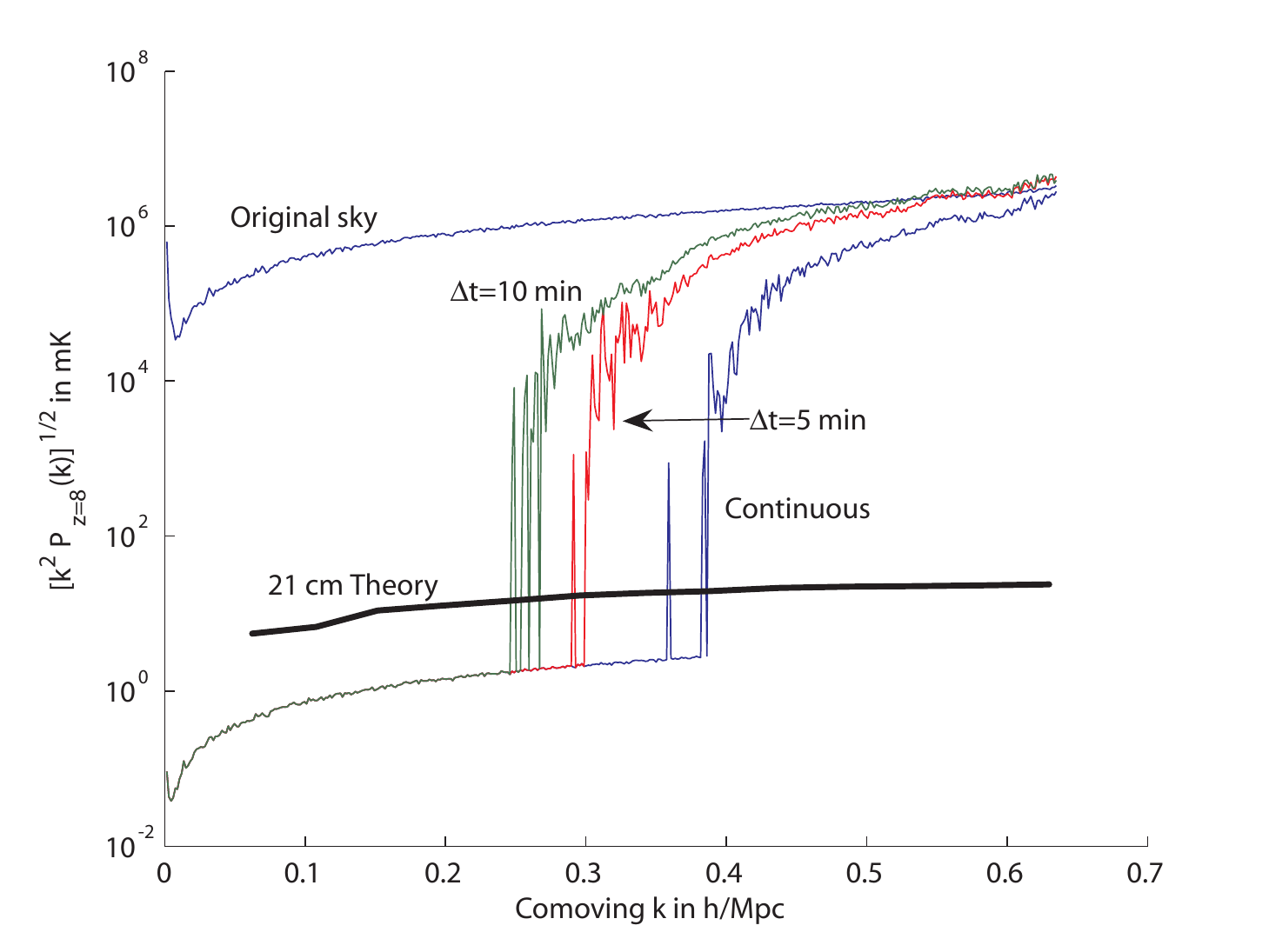}
\caption{Effect of temporal binning: 2D power spectra for the fiducial model but with varying binning time $\Delta t$, showing that one should strive for continuous $u$-$v$ coverage.}
\label{deltat}
\end{figure}

\subsubsection{Rotation Synthesis}
\label{rotsynth}
In a maximally optimistic situation, rotation synthesis can dramatically boost one's ability to image the sky, since for every baseline, one effectively obtains an entire arc of baselines on the $u$-$v$ plane.  However, rotation synthesis is not always as readily available as one might hope.  For instance, a declination zero object as viewed from the equator does not rotate but merely translates across the sky, allowing  no rotation synthesis.

As another example, consider a patch of the sky that lies close to the horizon.  The short time interval between the rising and setting of the patch means that even if continuous rotation synthesis is possible, one may not obtain sufficient total rotation synthesis time.  (Observing the patch again on the next night does not alleviate the problem, for a sidereal day later one is simply sampling the same points on the $u$-$v$ plane again).  One can also imagine a situation where $u$-$v$ coverage is poor even for a patch that remains in the sky for most of the night because of limitations in hardware calibration and data flow.  In such a scenario, each baseline would sweep out a long track on the $u$-$v$ plane, but this track would only be sparsely sampled by a series of snapshots of the sky.  We thus use two separate parameters to parameterize the quality of rotation synthesis: the total rotation synthesis time and the time $\Delta t$ between snapshots of the sky.

As expected, the performance of our foreground subtraction algorithm improves as one increases the total rotation synthesis time.  In figure \ref{totalrotsynth}, we show the effect of allowing integration time to increase from a single snapshot to $6$ hours (beyond which there is no significant improvement in foreground subtraction).  Each curve was simulated by assuming continuous rotation synthesis (i.e. $\Delta t \rightarrow 0$) using an instrument located at latitude and longitude $(\lambda, \phi) = (-27^\circ,0^\circ)$ for a field with right ascension and declination 
$(\alpha, \delta) = (60^\circ, -30^\circ)$\footnote{The latitude, right ascension, and declination were chosen to match planned observations by the MWA.  The longitude was set to zero for computational simplicity since its precise value does not result in qualitative changes in rotation synthesis.}.  We see that the first few hours of integration give rise to dramatic improvements, and that the gains begin to saturate after about $4\,\textrm{hours}$.  Intuitively, the saturation occurs because after several hours of integration, one is revisiting parts of the plane that have already been well sampled by other baselines, and so there is no further improvement\footnote{That does of course not imply that one should stop integrating after $4\,\textrm{hours}$ of observation, for repeated measurements of the same $u$-$v$ points increases signal-to-noise.  Indeed, current observation plans call for thousands of hours of integration.}.  As mentioned earlier in this section, the effectiveness of rotation synthesis depends on the location of the array as well as the sky coordinates of the patch being observed.  We find, however, that our conclusion is fairly robust to changes in array location and sky coordinates.  In other words, after 4 hours of observation one has essentially exhausted the potential of rotation synthesis, however small or large this potential may be.

In figure \ref{deltat}, we instead fix the integration time at $6$ hours and vary $\Delta t$.  It is clear that one should strive for continuous $u$-$v$ coverage if possible.  It should also be noted that with both parameters, it is generally the smallest scales that benefit from rotation synthesis -- the $r^{-2}$ array layout provides enough short baselines to ensure reasonable coverage on large angular scales even without exploiting Earth rotation.

\begin{figure}
\centering
\includegraphics[width=0.5\textwidth]{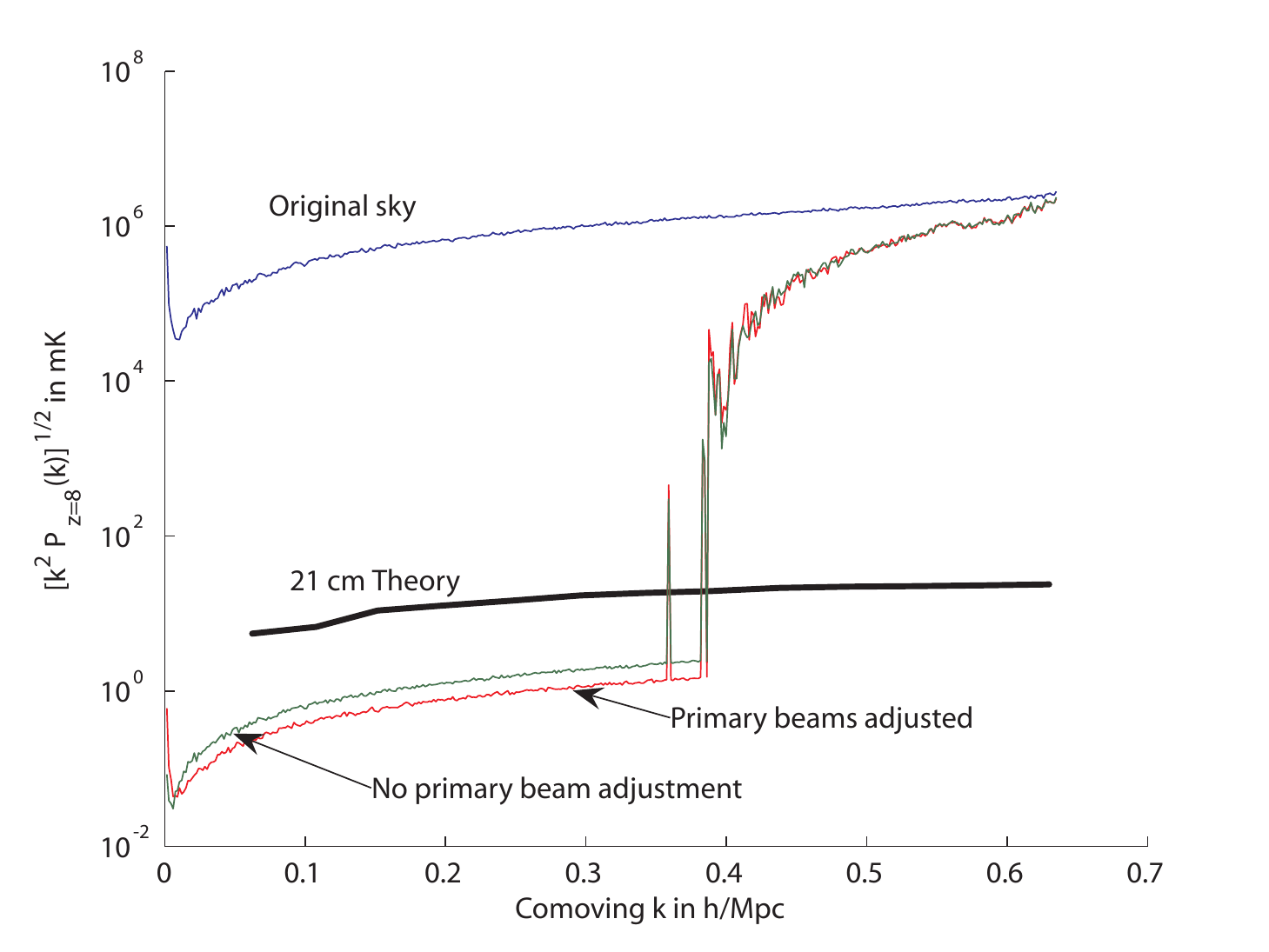}
\caption{Effect of primary beam equalization: 2D power spectra for the fiducial model but with different algorithms for dealing with the fact that the primary beam width changes with frequency.  Adjusting for the frequency dependence is seen to improve foreground subtraction slightly.}
\label{primary}
\end{figure}

\subsubsection{Primary beam width adjustments}
 
As mentioned in Section \ref{simstep}, the width of a radio array's primary beam is proportional to $\lambda$.  Thus, even if one has perfect coverage of the $u$-$v$ plane, the sky will look different at different frequencies.  In doing foreground subtraction, we consider two possible ways to analyze the data:
\begin{enumerate}
\item Do nothing, and simply accept the fact that the primary beam width changes with frequency.
\item Smear the $u$-$v$ plane data to smooth out the primary beams so that all the primary beams have the same width as the widest beam in the frequency range.
\end{enumerate}
The results are shown in figure \ref{primary}.  A close look at the plot reveals that adjusting for the frequency dependence of the primary beam does in fact bring about an improvement, albeit a small one, in our ability to subtract foregrounds.

\begin{figure}
\centering
\includegraphics[width=0.5\textwidth]{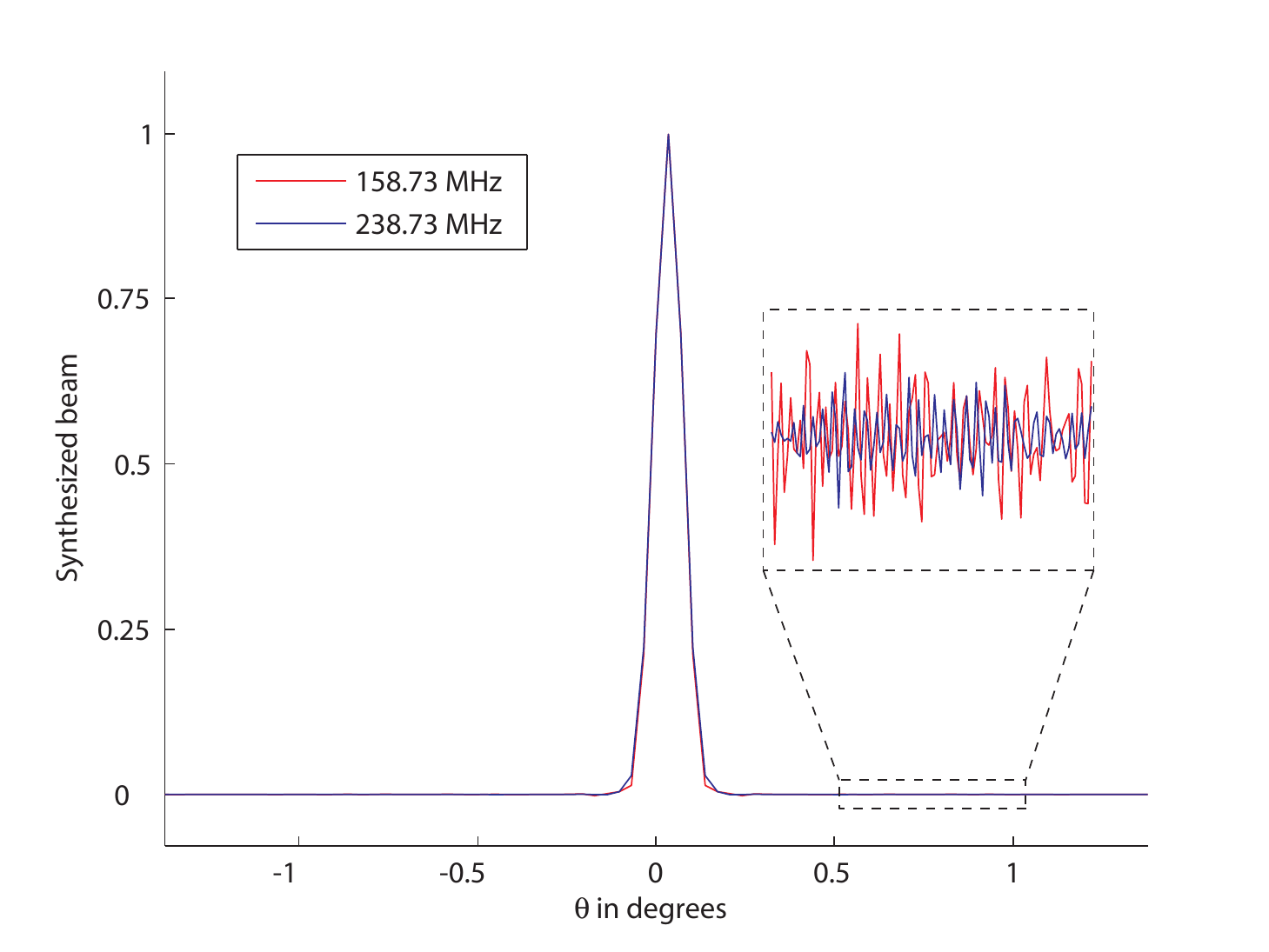}
\caption{Beam profiles after an extra Gaussian convolution, designed to make the heights and widths of the central peaks frequency-independent.  Parts of the beam beyond the central peak, however, will in general remain dependent on frequency.}
\label{adjwidth}
\end{figure}

\begin{figure}
\centering
\includegraphics[width=0.5\textwidth]{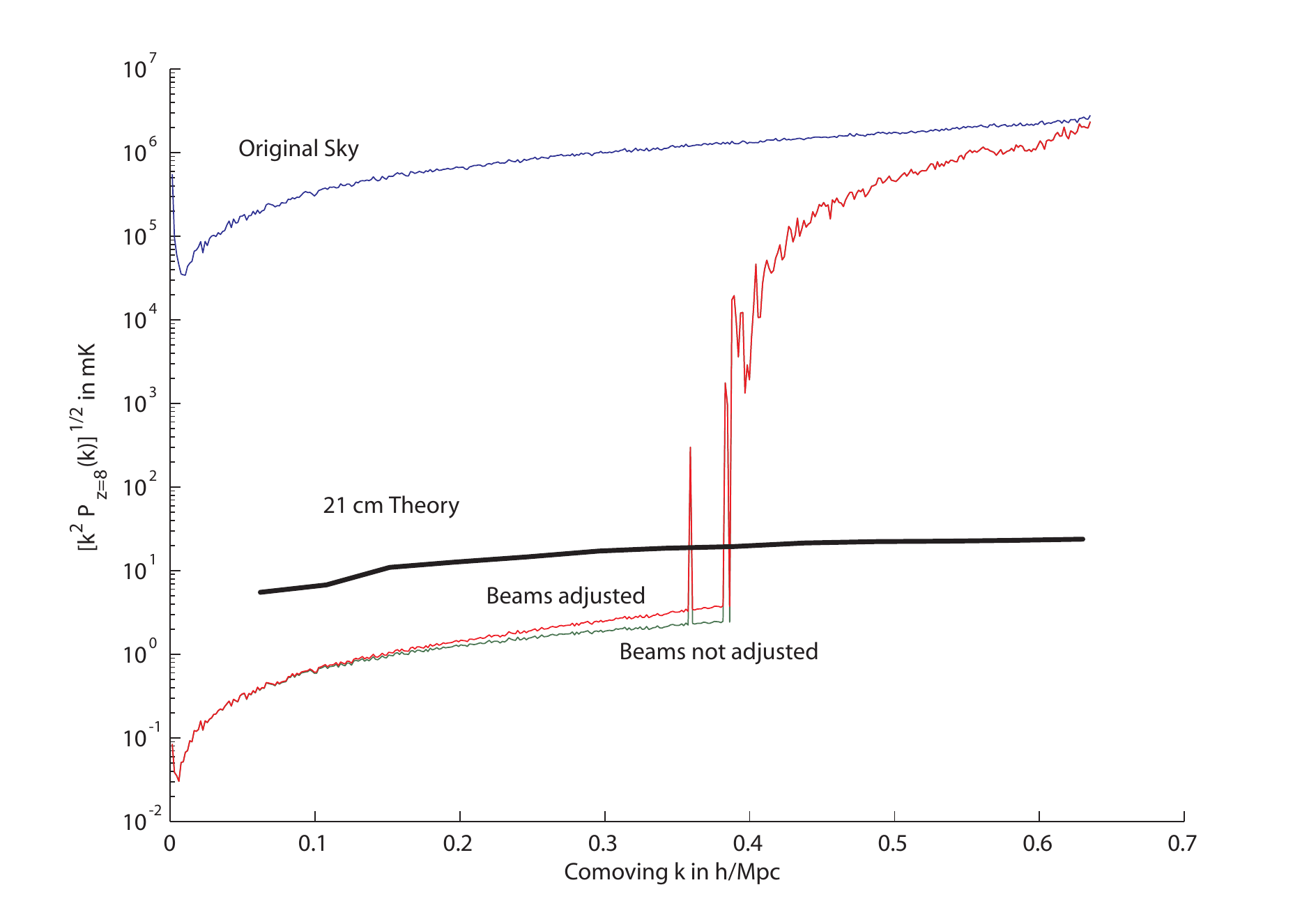}
\caption{Effect of synthesized beam adjustment: the attempt to smooth all maps to a common resolution before cleaning is seen to do more harm than good.}
\label{extraconv}
\end{figure}

\subsubsection{synthesized beam width adjustments}
\label{synthbeamadj}

The synthesized beam width is also expected to scale as $\lambda$, as illustrated in figure \ref{awesome}. 
In addition to the ``frizz'' issue that we have focused on so far, a second potential cause for concern is the width change of the central peak, as it means that sources slightly off from the centre of the beam will appear dimmer at higher frequency. This can potentially degrade the spectral fits. One way of adjusting for this is to convolve the dirty maps with frequency-dependent kernels whose frequency dependence exactly compensates for the changing width of the central part of the synthesized beam. The effect of this extra step is to ensure that, aside from the frizz effect, the sky has been convolved with the exact same beam at all frequencies. In other words, the angular resolution is made frequency independent by degrading the resolution of all maps to that of the lowest frequency map.  In figure \ref{adjwidth}, we show the profiles of the beams from figure \ref{awesome} after convolution with Gaussians of appropriate widths.

As mentioned above, the procedure tested in this section deals with the central peak, and has no mitigating effect on the ``frizz".  Since the ``frizz" is responsible for non-cosmological line-of-sight structure, one expects the beam adjustment to have very little effect in the frequency direction.  Figure \ref{extraconv} shows how this beam adjustment affects the the foreground subtraction in the spatial directions. 
On the largest and smallest scales, the extra convolution is seen to have no effect, and on intermediate scales smaller scales it is seen to make things slightly worse. 
This procedure of adjusting for changes in angular resolution with frequency therefore does more harm than good.
The harm presumably comes from the exacerbating the frizz-related problems (by causing departures from uniform 
weighting in the Fourier plane discussed in Section~\ref{weights} below).
The good comes from correcting for the above-mentioned dimming effect.
However, since $\lambda$ and hence the angular resolution varies by only a couple of percent across our frequency band, the point dimming effect will be very small for most point sources. Moreover, it will be a smooth function of frequency which can be accurately matched by our fitting polynomial, thus making angular resolution adjustments rather redundant.

\subsubsection{Flux Cuts}
\label{fluxcuts}

\begin{figure}
\centering
\includegraphics[width=0.5\textwidth]{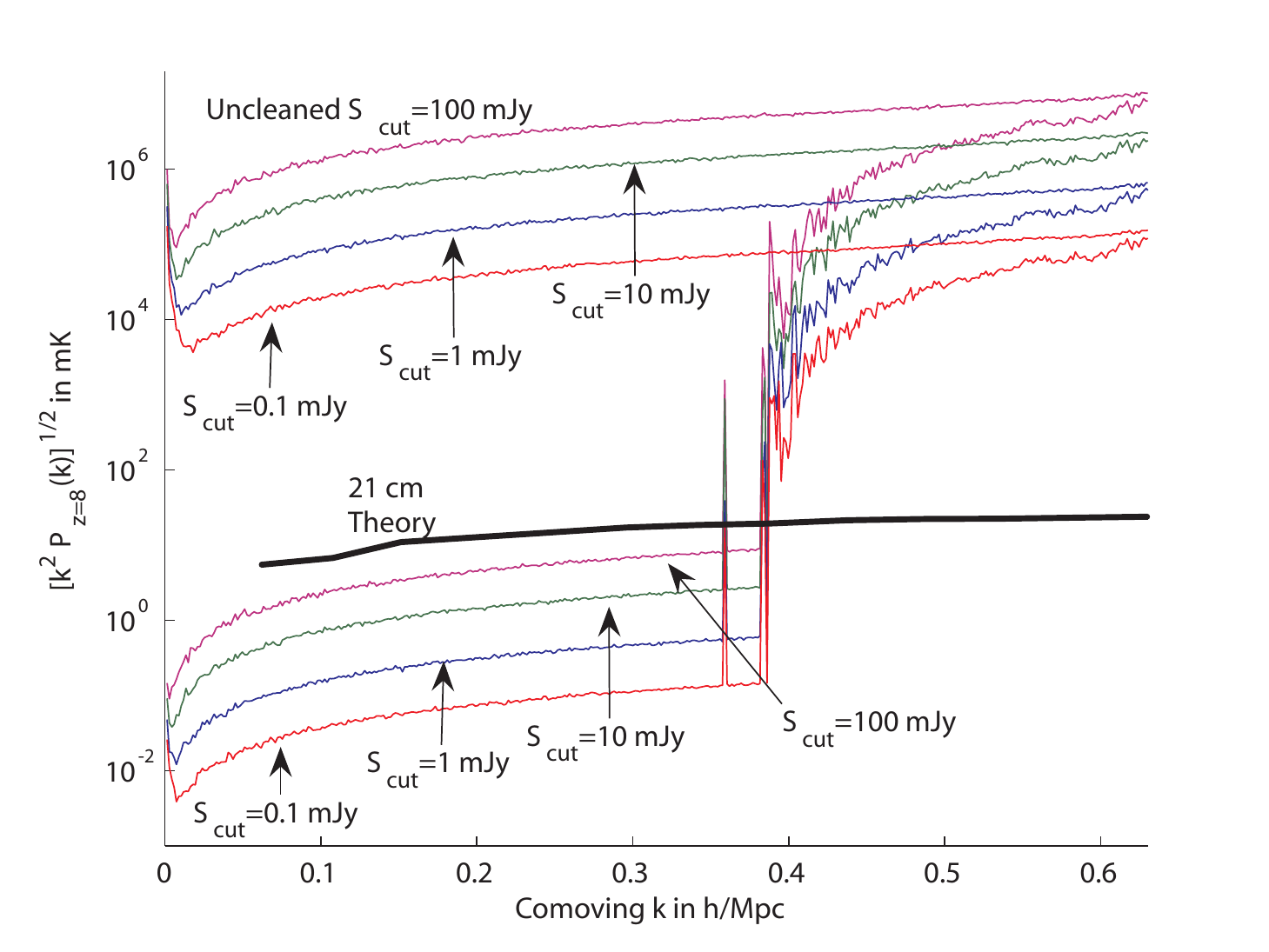}
\caption{Effect of flux cut for bright source removal: 2D power spectra for the fiducial model but with various bright point source flux cuts.}
\label{fluxcut}
\end{figure}

Since it remains unclear down to what flux cut $S_{\rm cut}$ one will be able to resolve and remove point sources with upcoming 21~cm tomography experiments, we examine how varying $S_{\rm cut}$ affects our algorithm for cleaning out unresolved point sources. In figure \ref{fluxcut}, we vary $S_{cut}$ from $0.1\,\textrm{mJy}$ to $100.0\,\textrm{mJy}$.
The results are shown in Figure~\ref{fluxcut}, and reveal a simple and useful scaling: 
raising the input power spectrum by some factor by increasing the flux cut raises the output by the same factor. 
This means that there is no need to rerun simulations with many flux cut levels, as the results from a single simulation can be analytically
scaled to apply to all other cases. All we need to extract from simulations is the factor by which the cleaning algorithm suppresses the point source fluctuations --- in this case, the suppression is seen to be about six orders of magnitude on large scales.

Quantitatively, the residuals are seen to lie below the theoretical 21-cm signal as long as the flux cut is below the $100.0\,\textrm{mJy}$ level.  However, there is some variability in results with frequency, both from variations in the residual power spectrum and in the magnitude of the theoretical curve. To be conservative, it is therefore prudent to aim to be able to remove bright point sources down to the $10\,\textrm{mJy}$ level.  This is consistent with the results in \citet{Judd08}, where it was found that foreground contaminants could be subtracted to an acceptable level starting from a sky model with a flux cut of $10\,\textrm{mJy}$.

These results were derived using quadratic fits. In section \ref{fitordersection}, we will see that the order of the polynomial has a strong effect on residuals, so if this $10$~mJy goal cannot be met, then the residual power spectrum can alternatively be brought down further by increasing the order of the fitting polynomial.

\begin{figure}
\centering
\includegraphics[width=0.5\textwidth]{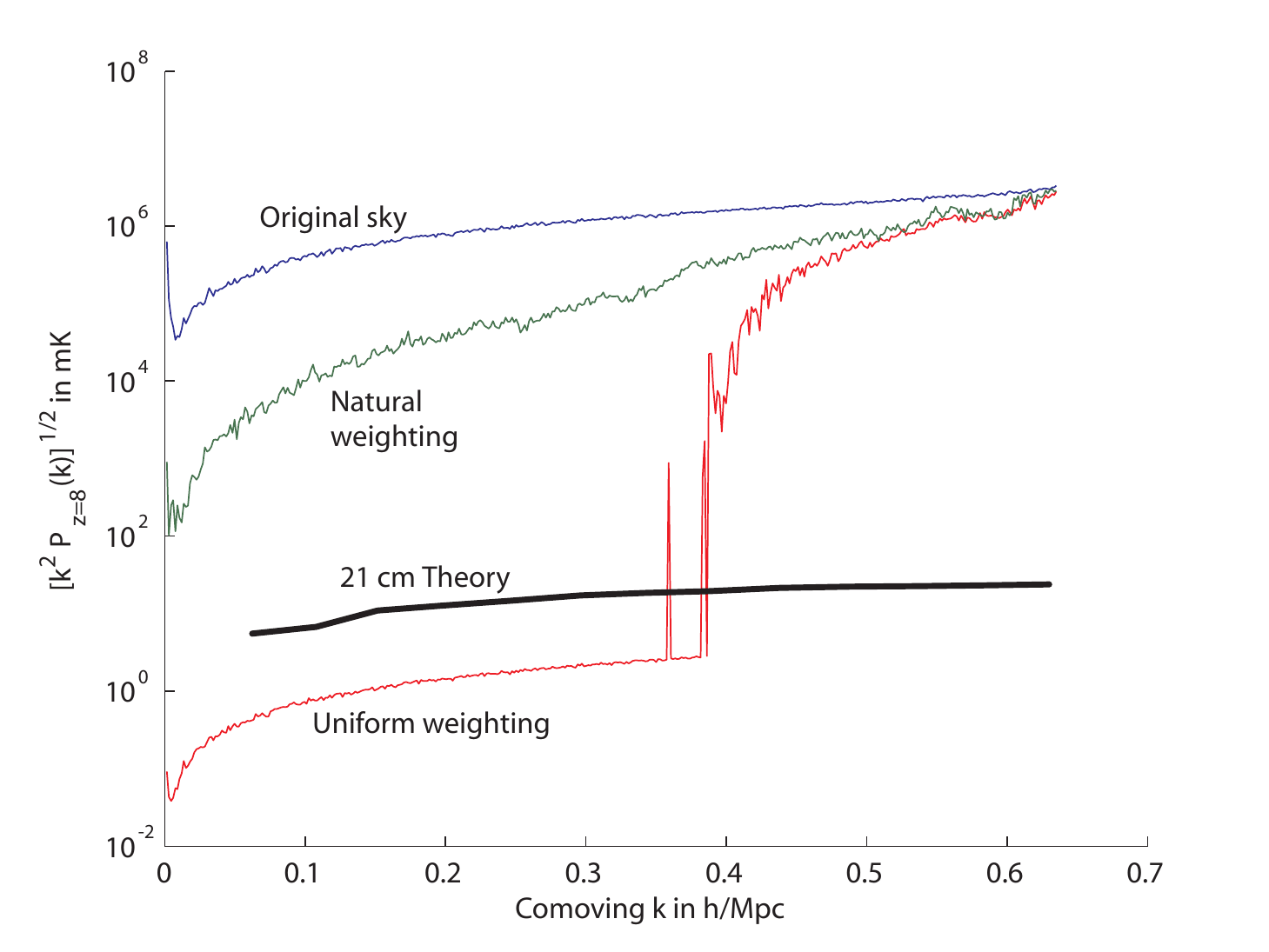}
\caption{Effect of $u$-$v$ plane weighting: 2D power spectra for the fiducial model but with two different weighting schemes for the $u$-$v$ plane. A uniform weighting of the $u$-$v$ plane is seen to far outperforms natural weighting.}
\label{weighting}
\end{figure}

\subsubsection{Weighting scheme on $u$-$v$ plane}
\label{weights}
In general, a radio array will not provide uniform coverage of the $u$-$v$ plane -- typically, some parts of the plane will be sampled more than once, while other parts will not be sampled at all.  To compensate for this, one may decide to weight different parts of the $u$-$v$ plane differently.  In figure \ref{weighting}, we examine a ``uniform" weighting scheme (where all sampled parts of the $u$-$v$ plane are given equal weight) as well as a ``natural" weighting scheme (where the $u$-$v$ measurements are not given any weighting beyond their ``natural" density as determined by the instrument).
It is clear that a uniform weighting far outperforms a natural one.  An intuitive explanation for why the uniform weighting so outperforms the natural weighting can be found in \citet{Judd08}.  Because the filling of the $u$-$v$ plane is accomplished slowly by a discrete set of baseline loci, the final distribution of baselines will not be particularly smooth.  The frequency dependence of the synthesized beam then maps this $u$-$v$ lumpiness into an incoherence in the frequency direction that is difficult to fit out using smooth polynomials.  With a uniform weighting, one artificially normalizes the baseline distribution so that each pixel in the sampled part of the $u$-$v$ plane has the same weight, thus ensuring that the distribution of baselines is smooth.

\begin{figure}
\centering
\includegraphics[width=0.5\textwidth]{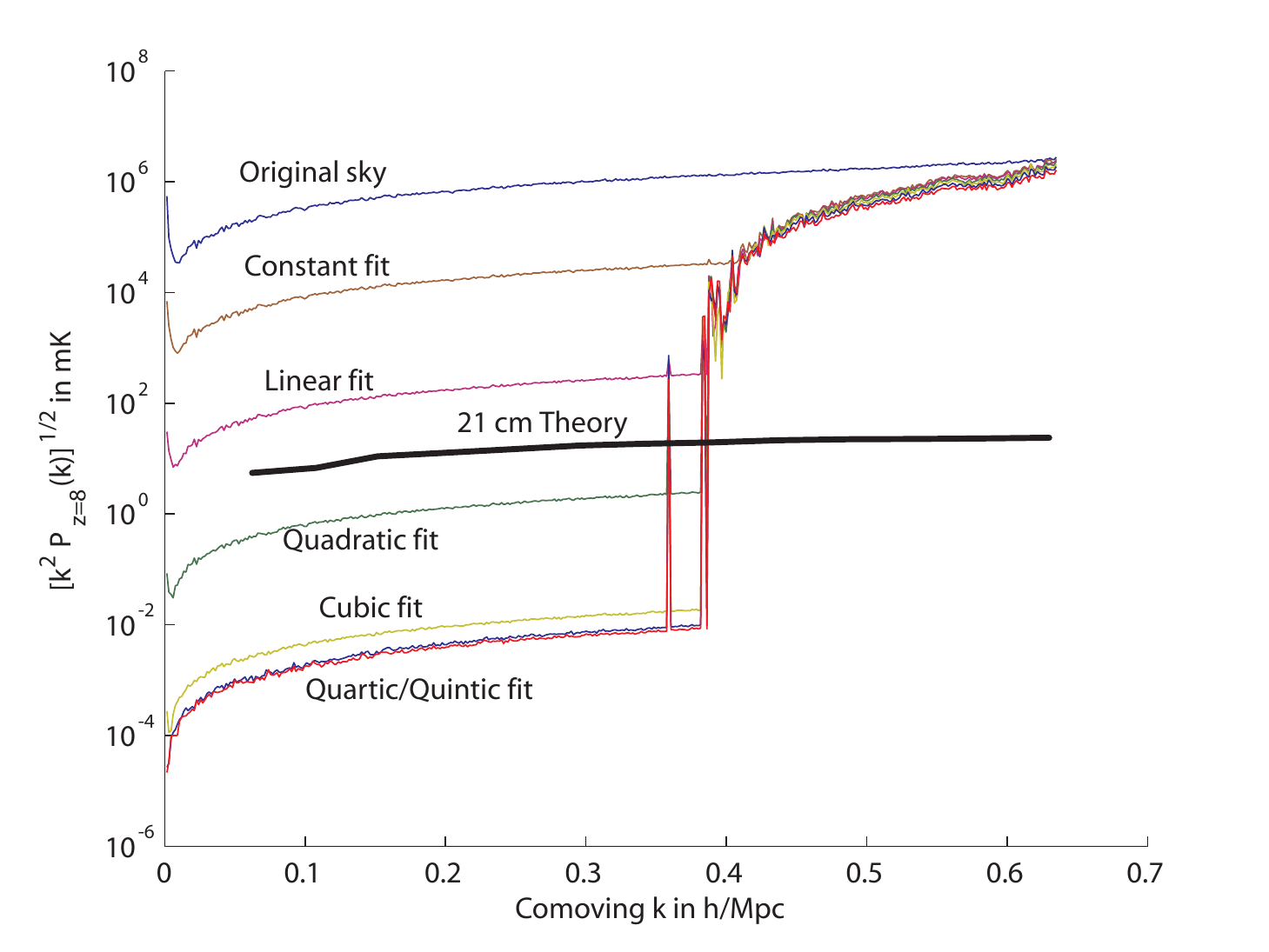}
\caption{Effect of fitting range: 2D power spectra for the fiducial model but with the foreground subtraction performed by fitting polynomials of various degrees to the spectra.}
\label{fitorder}
\end{figure}

\subsubsection{Order of the polynomial fit}
\label{fitordersection}
As we mentioned above, one should aim to fit the spectra with polynomials that are as low-order as possible, with the constraint that the residuals need to be below the expected cosmic signal level. Figure \ref{fitorder} shows that a quadratic fit satisfies this requirement for our fiducial case. If $S_{cut}$ cannot be pushed down to $10\,\textrm{mJy}$, however, then a higher order fit may be necessary. For example, 
$S_{cut}=100\,\textrm{mJy}$ would require a cubic fit.

\begin{figure}
\centering
\includegraphics[width=0.5\textwidth]{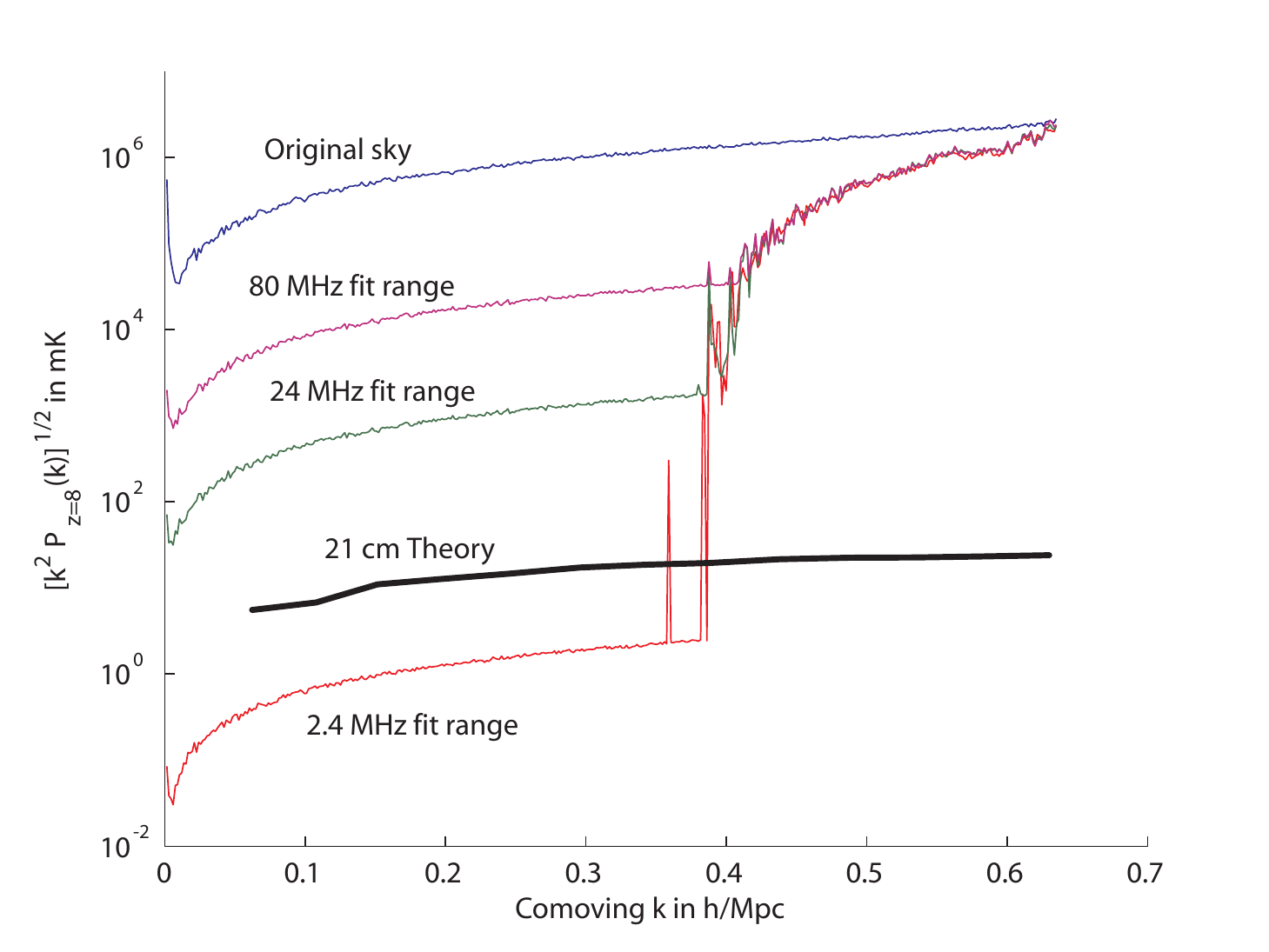}
\caption{Effect of fitting range: 2D power spectra for the fiducial model but with the foreground subtraction performed by fitting quadratic polynomials to the spectra over a variety of frequency ranges.}
\label{fitrange}
\end{figure}

\subsubsection{Range of the polynomial fit}
\label{rangeofpolyfit}

A choice must be made regarding the frequency range over which polynomial fits are applied to spectra of dirty map pixels. In general, one expects post-cleaning foreground residuals to increase with increasing frequency range, and this is what is seen in figure \ref{fitrange}, where the foreground cleaning was performed using quadratic fits over various frequency ranges.  
If quadratic polynomials are used, one should therefore not fit over the broad frequency range probed by a typical 21-cm tomography experiment.  Instead, one should divide the spectra into smaller frequency bands and subtract foregrounds individually from each band.

The effect of changing the frequency range is of course rather degenerate with the effect of changing the degree of the fitting polynomial, since what matters is ultimately the number of degrees of freedom that are fit out per unit frequency. As we increase either the polynomial order or the number of frequency bands, we are removing cosmological signal on ever smaller scales along the line of sight.
One should therefore fit over as broad a frequency range as possible subject to the constraint that the predicted residuals lie comfortably below the expected cosmic signal.  From figure \ref{fitrange}, we see that a frequency range of a few MHz appears appropriate for quadratic fits.

\section{Summary and Discussion}
\label{conc}

We have studied the problem of cleaning out point source foreground contamination from 21 cm tomography data, and investigated how the level of residual contamination after the cleaning process depends on various experimental and algorithmic parameters.
Our results show that while successful foreground removal is far from guaranteed, the signs are encouraging. For instance, the fact that the post-cleaning residual foregrounds lie below the simulated cosmological signal in the fiducial model of Section \ref{genresults} is promising, since the scenario in question is considered to be representative of current-generation experiments, and its cleaning algorithms are both simple and computationally cheap.
These conclusions agree with those of the independent and concurrent analysis of \citet{Judd08}, which is complementary by focusing on the specifics of the MWA experiment. As a further consistency check on our calculations and software, we and the authors of that paper agreed on a test example that we both simulated, obtaining consistent results.

We also identified a number of aspects of the problem that can be understood analytically. Noise and cosmic signal decouple from the point source problem as long as the cleaning method is linear. The power spectrum of unresolved point sources in the observed sky simply gets suppressed by a fixed $k$-dependent factor, so any shifts in this input power spectrum due to revised source count estimates or altered flux cuts for resolved source removal simply scale the output power spectrum by the same amount. Finally, simple Fourier space considerations explain why  line-of-sight cleaning works as well as it does.

Based on our analysis, we can make several specific recommendations regarding the foreground subtraction of unresolved point sources, pertaining to both instrumentation and data reduction.
\begin{itemize}
\item Instrumental recommendations:
\begin{itemize}
\item \textbf{Tile arrangement} should ensure good $u$-$v$ after rotation synthesis; we found both a monolithic arrangement and an $r^{-2}$ arrangement to work well with respect to foreground subtraction.
\item \textbf{Rotation synthesis} is important, but once an array has achieved $3$ to $4$ hours of integration, the advantage gained from further $u$-$v$ coverage is minimal. 
\item \textbf{Time between snapshots} taken of the sky should be made as short as possible.  Ideally, one should have continuous $u$-$v$ coverage, as is planned for experiments such as the MWA.
\end{itemize}
\item Data reduction/algorithmic recommendations:
\begin{itemize}
\item \textbf{Uniform weighting} of the $u$-$v$ plane outperforms ``natural weighting''.
\item \textbf{Adjusting for changes in the primary beam with frequency} through an extra convolution kernel in $u$-$v$ space results in a modest improvement in foreground subtraction.  However, this conclusion may change if the primary beam has sidelobes.
\item \textbf{Adjusting for changes in the synthesized beam with frequency} appears not to be worthwhile.
\item \textbf{Quadratic fits} across a few MHz are adequate for subtracting off spectrally smooth foreground components from the data
as long as bright point sources can be resolved and removed down to the $10\,\textrm{mJy}$ level --- otherwise higher orders should be considered.
\item \textbf{The frequency range} over which the polynomial fitting is performed should be narrow enough to allow one to see cosmic signal, but broad enough to prevent over-fitting which excessively removes cosmic signal. For quadratic fits, a range of a few MHz appears appropriate.
\item \textbf{The ability to remove bright point sources} prior to the subtraction of unresolved point sources is crucial, with the corresponding flux cut $S_{cut}$ being the most important parameter of all in our analysis.
If one can safely fit a cubic to the spectra without loosing too much cosmic signal, then one needs to be able to remove resolved points sources of flux down to $100\,\textrm{mJy}$.
\end{itemize}
\end{itemize}

Our results were deliberately conservative, obtained with an extremely simple line-of-sight cleaning procedure, and it is likely that one can do better. An interesting question for future work is determining the optimal procedure. For example, other classes of fitting functions may be worth considering, and it may be better to replace multiple polynomial fits in disjoint bands by a single fit to a function with more parameters, say a spline. 
Such work should quantify the extent to which one is removing power from the cosmological signal, and optimize the tradeoff between 
more foreground removal and less signal removal. Such an optimization should ideally be done using a more complete sky model, including diffuse Galactic synchrotron radiation.

For now, however, what is reassuring for the field of $21$-cm tomography is the fact that all of the recommendations listed above can be followed without any substantial revisions to current experimental designs. The papers mentioned in the introduction have already shown that neither noise nor diffuse Galactic foreground emission appear to pose insurmountable obstacles to 21~cm cosmology. 
Our results on point source cleaning suggest that arguably the most worrisome remaining foreground problem is also surmountable, allowing 21~cm cosmology to live up to its great potential.

\section*{Acknowledgments}
We  wish to thank Judd Bowman, Jacqueline Hewitt and Miguel Morales for helpful discussions and comments, and for sharing an early version of their manuscript prior to publication.  
We also thank Mike Matejek and Christopher Williams for useful feedback, and 
Matthew McQuinn for sharing 21~cm simulation results.
This work was supported by NSF grants AST-0134999 and AST-05-06556 as well as fellowships from the David and Lucile
Packard Foundation and the Research Corporation.  

\bibliographystyle{mn2e}
\bibliography{21cmps}

\end{document}